\documentclass[
 amsmath,
 amssymb,
 aps,
]{revtex4-2}

\usepackage{graphicx} % Include figure files
\usepackage{dcolumn} % Align table columns on decimal point
\usepackage{bm} % bold math

\usepackage{hyperref} % add hypertext capabilities

\usepackage[linesnumbered, ruled, vlined]{algorithm2e}

\usepackage{xcolor}

\usepackage{amsthm}

\usepackage{xr}
\usepackage{slashed}

\usepackage{tikz}
\usetikzlibrary{quantikz}

\allowdisplaybreaks

\makeatletter
\@addtoreset{equation}{subsection}
\makeatother

\usepackage{MnSymbol}

\usepackage{soul}

\makeatletter
\newcommand{\vast}{\bBigg@{3.0}}
\newcommand{\Vast}{\bBigg@{4.0}}

\makeatother

\begin{document}

% !TEX root = ./paper_population_dynamics_000_001_main.tex

\preprint{APS/123-QED}

\title{Steady-state thermodynamics for population dynamics in fluctuating environments with side information}
% \thanks{A footnote to the article title}

\author{Hideyuki Miyahara}
% \altaffiliation[Also at ]{Physics Department, XYZ University.}
\email{miyahara@g.ucla.edu, hmiyahara512@gmail.com}
% \email{miyahara@g.ucla.edu}
% \email{hmiyahara512@gmail.com}

\affiliation{
Department of Electrical and Computer Engineering, \\
Henry Samueli School of Engineering and Applied Science, \\
University of California, Los Angeles, California 90095
}

% \affiliation{
% International Research Center for Neurointelligence (WPI-IRCN), UTIAS,
% The University of Tokyo,
% % 7-3-1 Hongo, Bunkyo-ku,
% Tokyo 113_0033, Japan
% }

% \author{Vwani Roychowdhury}
%
% \email{vwani@g.ucla.edu}
%
% \affiliation{
% Department of Electrical and Computer Engineering,
% Henry Samueli School of Engineering and Applied Science,
% University of California, Los Angeles, California 90095
% }

\date{\today}

% !TEX root = ./paper_population_dynamics_000_001_main.tex

\begin{abstract}
  Steady-state thermodynamics (SST) is a relatively newly emerging subfield of physics, which deals with transitions between steady states.
  In this paper, we find an SST-like structure {in} population dynamics of organisms that can sense their fluctuating environments.
  As heat is divided into two parts in SST, we decompose population growth into two parts: housekeeping growth and excess growth.
  {Then, we derive the Clausius equality and inequality for excess growth.}
  {Using numerical simulations}, we demonstrate how the Clausius inequality behaves depending on the magnitude of noise and strategies that organisms employ.
  Finally, we discuss the novelty of our findings and compare them with a previous study.
\end{abstract}

% \keywords{Suggested keywords}

\maketitle

% \tableofcontents

% !TEX root = ./paper_population_dynamics_000_001_main.tex

\section{Introduction}

Thermodynamics is one of the most important subfields of physics and provides the foundation for other subfields in physics.
Furthermore, the last two decades have witnessed novel developments in thermodynamics: stochastic thermodynamics and steady-state thermodynamics (SST).
In stochastic thermodynamics, Jarzynski's equality and Crooks' theorem succeed in unifying the second law of thermodynamics and linear response theory~\cite{Jarzynski_001, Jarzynski_003, Crooks_001, Crooks_002, Seifert_001}, and they are generalized from the viewpoint of information theory~\cite{Sagawa_001, Sagawa_002, Sagawa_003, Sagawa_005, Parrondo_001, Miyahara_007}.
On the other hand, in SST, {which} deals with transitions between {non-equilibrium} steady states, {the Clausius equality and inequality for excess heat during transitions between non-equilibrium steady states} were found~\cite{Oono_001, Speck_001, Esposito_001, Chetrite_001, Sekimoto_003, Sasa_001, Komatsu_001, Komatsu_002, Sagawa_006}.

In theoretical biology, population dynamics, which involves the study of the dynamics of the number of individuals and organisms, is a particularly important topic.
In particular, population dynamics of organisms that switch their phenotypes stochastically has attracted much attention and has been investigated extensively because it is quite essential to understand how organisms adapt to and survive their fluctuating environments~\cite{Kussell_001, Leibler_001, Hermisson_001, Georgii_001, Baake_001, Baake_002, Sughiyama_001}.
Recently, population dynamics and thermodynamics have been integrated, and interdisciplinary research between population dynamics and thermodynamics has witnessed some novel findings.
For example, FRs for population dynamics were found~\cite{Cover_002, Kobayashi_003, Kobayashi_004, Miyahara_008}.
Furthermore, an SST-like structure was found for a biological system~\cite{Sughiyama_002, Levien_001, Kwon_001}.

In this paper, we construct an SST-like structure for population dynamics of organisms that can sense their environments.
As mentioned above, an SST-like structure for population dynamics was investigated in Ref.~\cite{Sughiyama_002}, but only a fixed transition matrix was considered.
Furthermore, this previous work assumed that organisms cannot sense their environment.
When we consider the adaption of organisms to their environments, the ability of sensing environments is critical and must be taken into account.
Then, we extend the work by introducing information on environments and derive new relations for population dynamics.

This paper is organized as follows.
In Sec.~\ref{sec_prerequisites_001}, we explain a system of interest and define some important quantities.
In Sec.~\ref{sec_main_results_001}, we describe the main results of this paper.
In Secs.~\ref{sec_proof_first_main_001} and \ref{sec_proof_second_main_001}, we provide the proofs of the main results.
Then, in Sec.~\ref{sec_numerical_simulation_001}, we perform numerical simulations to confirm the findings.
{In Sec.~\ref{sec_discussions_001}, we discuss the main results by comparing them with a trivial bound and the finding in Ref.~\cite{Sughiyama_002}.}
Section~\ref{sec_conclusions_001} concludes this paper.

%

% !TEX root = ./paper_population_dynamics_000_001_main.tex

\section{Prerequisites} \label{sec_prerequisites_001}

In this section, we first define some variables that characterize population dynamics and some quantities that play {a critical} role in developing an SST-like structure for population dynamics.
We also define steady states and two types of population growth: housekeeping growth and excess growth.

\subsection{System variables}

We consider organisms with $n$ phenotypes and assume that phenotype-switching and duplication occur at {discrete time} $\tau=1, 2, \dots, t$.
To describe it mathematically, let us introduce three variables: $N_\tau (x)$, $y_\tau$, and $z_\tau$ for $x = 1, 2, \dots, n$ and $\tau = 1, 2, \dots, t$.
Here $N_\tau (x)$ is the number of organisms of phenotype $x$ at time $\tau$, $y_\tau$ is the state of the environment at time $\tau$, and $z_\tau$ represents information on $y_\tau$.
Furthermore, we assume that organisms do not know $y_\tau$ but may utilize $z_\tau$ as information on $y_\tau$.
In addition, we impose the following relations: $N_\tau (x) \ge 0$ for $x = 1, 2, \dots, n$ and $\tau = 1, 2, \dots, t$.
We also define $\vec{N}_\tau \coloneqq [N_\tau (1), N_\tau (2), \dots, N_\tau (n)]^\intercal$.

\subsection{Dynamics and detailed balance condition}

We now describe the system of interest that is composed of two terms: phenotype-switching and population growth.
We provide two mathematical expressions of the target system: the stochastic equation and its matrix form since they are both helpful.

\subsubsection{Dynamics in the stochastic equation} \label{sec_dynamics_stochastic_001}

In this paper, we consider the system governed by
\begin{align}
N_\tau (x) &= \mu_{y_\tau} (x) \sum_{x' = 1}^n T_{z_\tau} (x| x') N_{\tau - 1} (x'), \label{eq_dynamics_X_discrete_001_001}
\end{align}
where $T_z (x| x')$ represents {the phenotype-switching rate} from phenotype $x'$ to phenotype $x$ in the case of $z$ and {$\mu_y (x)$} is {the duplication rate of organisms whose phenotype is $x$} when the state of the environment is {$y$}.
Thus, $T_z (\cdot | \cdot)$ satisfies $\sum_{x = 1}^n T_z (x| x') = 1$.
Each organism changes its phenotype by $T_z (\cdot | \cdot)$ in Eq.~\eqref{eq_dynamics_X_discrete_001_001} typically to maximize its total population.

We assume that $T_z (\cdot | \cdot)$ satisfies the detailed balance condition
\begin{align}
T_z (x| x') w_z^T (x') &= T_z (x'| x) w_z^T (x), \label{eq_detailed_balance_condition_001_001}
\end{align}
where $w_z^T (x)$ is the stationary distribution induced by $T_z (\cdot | \cdot)$.
The detailed balance condition is originally an assumption widely used in thermodynamics; otherwise, we cannot define thermodynamic quantities of the system.
We also define
\begin{align}
H_{y, z} (x| x') \coloneqq \mu_y (x) T_z (x| x'), \label{eq_def_H_001_001}
\end{align}
for $x, x' = 1, 2, \dots, n$.

\subsubsection{Dynamics in the matrix form}

We provide a matrix expression of the dynamics defined in Sec.~\ref{sec_dynamics_stochastic_001}.
That is, we rewrite Eq.~\eqref{eq_dynamics_X_discrete_001_001} as
\begin{align}
\vec{N}_t &= \hat{M}_{y_t} \hat{T}_{z_t} \vec{N}_{t-1}, \label{eq_dynamics_X_discrete_001_002}
\end{align}
where $\hat{M}_y$ and $\hat{T}_z$ are the $n \times n$ matrices given, respectively, by
\begin{align}
\hat{M}_y &\coloneqq \mathrm{diag} [\mu_y (1), \mu_y (2), \dots, \mu_y (n)], \\
\hat{T}_z &\coloneqq
\begin{bmatrix}
T_z (1| 1) & T_z (1| 2) & \dots & T_z (1| n) \\
T_z (2| 1) & T_z (2| 2) & \dots & T_z (2| n) \\
\vdots & \vdots & \ddots & \vdots \\
T_z (n| 1) & T_z (n| 2) & \dots & T_z (n| n)
\end{bmatrix}.
\end{align}
From Eq.~\eqref{eq_dynamics_X_discrete_001_002}, we also have
\begin{align}
\vec{N}_t &= \bigg[ \prod_{\tau=1}^t \hat{M}_{y_\tau} \hat{T}_{z_\tau} \bigg] \vec{N}_0.
\end{align}

Similarly, the matrix representation of Eq.~\eqref{eq_def_H_001_001} is also introduced as
\begin{align}
\hat{H}_{y, z} &\coloneqq \hat{M}_y \hat{T}_z.
\end{align}
Note that $[\hat{H}_{y, z}]_{x, x'} = H_{y, z} (x| x')$ where $[\cdot]_{x, x'}$ is an element in the $x$-th row and the $x'$-th column of the matrix.

\subsection{Population growth}

{Population growth} is of great importance in population dynamics; we here introduce it by
\begin{align}
\Phi_{0 \to t} \coloneqq& \ln \frac{N_t}{N_0}, \label{eq_population_growth_rate_001_001}
\end{align}
where $N_\tau \coloneqq \sum_{x=1}^n N_\tau (x)$ is the total number of organisms at time $\tau$.
We also define $\phi_\tau \coloneqq \ln \frac{N_\tau}{N_{\tau - 1}}$; so we have $\Phi_{0 \to t} = \sum_{\tau=0}^t \phi_\tau$.

\subsection{Steady states}

We introduce steady states, which play a critical role in SST.
When $y_\tau$ and $z_\tau$ are fixed or vary sufficiently slowly, $\phi_\tau$ reaches a steady state labeled by $y_\tau$ and $z_\tau$.
The above situation corresponds to a quasi-static process in thermodynamics, and thus it is of great importance.

Let us denote the normalized steady state by $\vec{v}_{y, z}^\mathrm{st} \coloneqq [v_{y, z}^\mathrm{st} (1), v_{y, z}^\mathrm{st} (2), \dots, v_{y, z}^\mathrm{st} (n)]^\intercal$.
In the case of $y_\tau = y$ and $z_\tau = z$, $\vec{v}_{y, z}^\mathrm{st}$ satisfies
\begin{align}
v_{y, z}^\mathrm{st} (x) &= e^{- \phi_{y, z}^\mathrm{st}} \mu_y (x) \sum_{x' = 1}^n T_z (x| x') v_{y, z}^\mathrm{st} (x'), \label{eq_steady_state_001_001}
\end{align}
where $\phi_{y, z}^\mathrm{st}$ is $\phi_\tau$ in {the case of} the steady state labeled by $y$ and $z$.
Note that $\vec{v}_{y, z}^\mathrm{st}$ is normalized such that $\| \vec{v}_{y, z}^\mathrm{st} \|_1 = 1$, while $\vec{N}_\tau$ is not; thus, at a steady state, we have {$\vec{v}_{y_\tau, z_\tau}^\mathrm{st} = \vec{N}_\tau / N_\tau$}.
Mathematically, {steady-state population growth} per time step is given by
\begin{align}
\phi_{y, z}^\mathrm{st} &= \ln \sum_{x, x' = 1}^n u_{y, z}^\mathrm{st} (x) \mu_y (x) T_z (x| x') v_{y, z}^\mathrm{st} (x'),
\end{align}
where $\vec{u}_{y, z}^\mathrm{st} \coloneqq \big[ u_{y, z}^\mathrm{st} (1), u_{y, z}^\mathrm{st} (2), \dots, u_{y, z}^\mathrm{st} (n) \big]$ is the dual eigenvector of $\vec{v}_{y, z}^\mathrm{st}$ that satisfies {$\vec{u}_{y, z}^\mathrm{st} \vec{v}_{y, z}^\mathrm{st} = \sum_{x = 1}^n u_{y, z}^\mathrm{st} (x) v_{y, z}^\mathrm{st} (x) = 1$}.

\subsection{Eigenstates}

We define the left eigenvectors $\{ \vec{u}_{y, z}^\mathrm{eig} (x) \}_{x=1}^n$ and right eigenvectors $\{ \vec{v}_{y, z}^\mathrm{eig} (x) \}_{x=1}^n$ of $\ln \hat{H}_{y, z}$.
Then, $\{ \vec{u}_{y, z}^\mathrm{eig} (x) \}_{x=1}^n$ and $\{ \vec{v}_{y, z}^\mathrm{eig} (x) \}_{x=1}^n$ of $\ln \hat{H}_{y, z}$ satisfy, respectively,
\begin{align}
  \vec{u}_{y, z}^\mathrm{eig} (x) \ln \hat{H}_{y, z} &= \phi_{y, z}^\mathrm{eig} (x) \vec{u}_{y, z}^\mathrm{eig} (x), \\
  \ln \hat{H}_{y, z} \vec{v}_{y, z}^\mathrm{eig} (x) &= \phi_{y, z}^\mathrm{eig} (x) \vec{v}_{y, z}^\mathrm{eig} (x),
\end{align}
where the elements of $\vec{v}_{y, z}^\mathrm{eig} (x)$ and $\vec{u}_{y, z}^\mathrm{eig} (x)$ are nonnegative, $\| \vec{v}_{y, z}^\mathrm{eig} (x) \|_1 = 1$ for $x = 1, 2, \dots, n$, and $\{ \phi_{y, z}^\mathrm{eig} (x) \}_{x=1}^n$ are the eigenvalues of $\ln \hat{H}_{y, z}$ that satisfy $\phi_{y, z}^\mathrm{eig} (1) \ge \phi_{y, z}^\mathrm{eig} (2) \ge \dots \ge \phi_{y, z}^\mathrm{eig} (n-1) \ge \phi_{y, z}^\mathrm{eig} (n)$.
Furthermore, we impose
\begin{align}
\sum_{x = 1}^n \vec{v}_{y, z}^\mathrm{eig} (x) \vec{u}_{y, z}^\mathrm{eig} (x) &= \hat{1}_n,
\end{align}
for the normalization of $\vec{u}_{y, z}^\mathrm{eig} (x)$ for $x = 1, 2, \dots, n$.
Here $\hat{1}_n$ is the $n \times n$ identity matrix.
By definition, we have $\vec{v}_{y, z}^\mathrm{st} = \vec{v}_{y, z}^\mathrm{eig} (1)$, $\vec{u}_{y, z}^\mathrm{st} = \vec{u}_{y, z}^\mathrm{eig} (1)$, and $\phi_{y, z}^\mathrm{st} = \phi_{y, z}^\mathrm{eig} (1)$.

\subsection{Housekeeping growth and excess growth}

In SST, heat is divided into two parts, that is, housekeeping and excess parts.
Similarly, we divide {population growth} into two parts:
\begin{align}
\Phi_{0 \to t} &= \Phi_{0 \to t}^\mathrm{hk} + \Phi_{0 \to t}^\mathrm{ex}, \label{eq_population_growth_decomposition_001_001}
\end{align}
where {housekeeping growth} and excess growth are given, respectively, by
\begin{align}
\Phi_{0 \to t}^\mathrm{hk} &\coloneqq \sum_{\tau = 1}^t \phi_{y_\tau, z_\tau}^\mathrm{st}, \label{eq_def_housekeeping_growth_001_001} \\
\Phi_{0 \to t}^\mathrm{ex} &\coloneqq \Phi_{0 \to t} - \Phi_{0 \to t}^\mathrm{hk}. \label{eq_def_excess_growth_001_001}
\end{align}
Thus, the main purpose of this paper is to consider the Clausius equality and inequality for {excess growth, Eq.~\eqref{eq_def_excess_growth_001_001}}.

\subsection{Forward and backward path probabilities}

Path probabilities are useful for the derivation of the main results of this paper.
First we define the forward path probability as follows:
\begin{align}
p_\mathrm{f} (X_{0 \to t}| Z_{0 \to t}) &\coloneqq \bigg[ \prod_{\tau=1}^t T_{z_\tau} (x_\tau| x_{\tau - 1}) \bigg] p_\mathrm{ini} (x_0),
\end{align}
where $X_{0 \to t} \coloneqq \{ x_\tau \}_{\tau = 0}^t$ and $Z_{0 \to t} \coloneqq \{ z_\tau \}_{\tau = 0}^t$.
We also define the backward path probability as
\begin{align}
p_\mathrm{b} (X_{0 \to t}| Y_{0 \to t}, Z_{0 \to t}) &\coloneqq \frac{\Big( \prod_{\tau=1}^t \mu_{y_\tau} (x_\tau) \Big) p_\mathrm{f} (X_{0 \to t}| Z_{0 \to t})}{\Big\langle \prod_{\tau=1}^t \mu_{y_\tau} (x_\tau) \Big\rangle_{p_\mathrm{f} (X_{0 \to t}| Z_{0 \to t})}}, \label{eq_backward_path_probability_003_001}
\end{align}
where $Y_{0 \to t} \coloneqq \{ y_\tau \}_{\tau = 0}^t$ and $\langle f (x) \rangle_{g (x)} \coloneqq \sum_{x = 1}^n f (x) g (x)$.
The backward path probability, Eq.~\eqref{eq_backward_path_probability_003_001}, is also expressed as
\begin{align}
p_\mathrm{b} (X_{0 \to t}| Y_{0 \to t}, Z_{0 \to t}) &= \bigg[ \prod_{\tau=1}^t \vec{u} (x_\tau) e^{\ln \hat{H}_{y_\tau, z_\tau} - \phi (\tau) \hat{1}_n} \vec{v} (x_{\tau-1}) \bigg] v_{y_0, z_0}^\mathrm{st} (x_0), \label{eq_backward_path_probability_003_002}
\end{align}
where
{\begin{align}
  \vec{u} (x) &\coloneqq [\underbrace{0, \dots, 0}_{x-1}, 1, \underbrace{0, \dots, 0}_{n-x}], \\
  \vec{v} (x) &\coloneqq [\underbrace{0, \dots, 0}_{x-1}, 1, \underbrace{0, \dots, 0}_{n-x}]^\intercal.
\end{align}}%
Our definitions of the forward and backward path probabilities are the same {as} those defined in Ref.~\cite{Kobayashi_003, Nozoe_001, Genthon_001}.
The forward path probability is the probability of observing an organism whose history is $X_{0 \to t}$ when we randomly pick out an organism at time $0$ and track its phenotype-switching in a time-forward manner.
{When the tracked organism duplicates,} we can randomly choose one of the daughters since the forward path probability does not depend on the choice.
On the other hand, the backward path probability is the probability of observing an organism of which history is $X_{0 \to t}$ when we randomly choose an organism at time $t$ and track it back in a time-backward manner.
We utilize Eq.~\eqref{eq_backward_path_probability_003_001} to characterize the necessary condition for the second main result of this paper.

% !TEX root = ./paper_population_dynamics_000_001_main.tex

\section{Main results} \label{sec_main_results_001}

This section is the main part of this paper, in which we state the main claims.
The first one is the Clausius equality {for excess growth, Eq.~\eqref{eq_def_excess_growth_001_001},} and the second one is the Clausius inequality {for excess growth, Eq.~\eqref{eq_def_excess_growth_001_001}}.

\subsection{First main result}

We state the first main result of this paper: the Clausius equality {for excess growth, Eq.~\eqref{eq_def_excess_growth_001_001}}.
Let us consider a quasi-static process with Eq.~\eqref{eq_dynamics_X_discrete_001_001}; that is, we focus on the process in which the initial state is given by $\vec{N}_0 = N_0 \vec{v}_{y_0, z_0}^\mathrm{st}$ and $\Delta y_\tau \coloneqq y_\tau - y_{\tau-1}$ and $\Delta z_\tau \coloneqq z_\tau - z_{\tau-1}$ are sufficiently small {($| \Delta y |, | \Delta z| \ll 1$)} such that $\vec{N}_\tau / N_\tau = \vec{v}_{y_\tau, z_\tau}^\mathrm{st}$ for $\tau = 0, 1, 2, \dots, t$.

If $w_z^T (x)$ in Eq.~\eqref{eq_detailed_balance_condition_001_001} is independent of $z$ for $x = 1, 2, \dots, n$:
\begin{align}
  \nabla_{z} w_z^T (x) &= 0, \label{eq_condition_w_001_001}
\end{align}
where $\nabla_z$ is the derivative with respect to $z$, we obtain the Clausius equality {for excess growth, Eq.~\eqref{eq_def_excess_growth_001_001}}:
\begin{align}
  \Phi_{0 \to t}^\mathrm{ex} &= S (y_t, z_t) - S (y_0, z_0), \label{eq_Clausius_equality_001_001}
\end{align}
where the pseudo-entropy $S (y, z)$ is given by~\footnote{The pseudo-entropy~\eqref{eq_def_pseudo-entropy_001_001} is not the entropy, but in our framework, it plays an important role similar to the entropy in thermodynamics; so we call it the pseudo-entropy.}
\begin{align}
  S (y, z) &\coloneqq \frac{1}{2} \ln \sum_{x = 1}^n u_{y, z}^\mathrm{st} (x) w_z^T (x). \label{eq_def_pseudo-entropy_001_001}
\end{align}

Here, we provide an example in which Eq.~\eqref{eq_condition_w_001_001} is satisfied.
Let us consider a system of two phenotypes.
If we set
\begin{align}
  \hat{T}_{z} &=
  \begin{bmatrix}
    1 - \omega_{z} & \omega_{z} \\
    \omega_{z} & 1 - \omega_{z}
  \end{bmatrix}, \label{eq_example_T_001_001}
\end{align}
where $[\hat{T}_{z}]_{x, x'} \coloneqq T_{z} (x| x')$, then the stationary distribution is given by $[w_{z}^T (1), w_{z}^T (2)] = [1/2, 1/2]$ for any $\omega_{z}$; thus, Eq.~\eqref{eq_condition_w_001_001} is satisfied.
Equation~\eqref{eq_example_T_001_001} can be easily generalized to the case of $n$ phenotypes.

We have assumed the quasi-static process for the above theorem.
Otherwise, this theorem does not hold since the eigenstate associated with the largest eigenvalue is not attainable.
The following theorem, however, holds even when the process is not quasi-static.

\subsection{Second main result}

Then we describe the second main result of this paper, that is, the Clausius inequality {for excess growth, Eq.~\eqref{eq_def_excess_growth_001_001}}.
Let us assume that the initial state is given by $\vec{N}_0 = N_0 \vec{v}_{y_0, z_0}^\mathrm{st}$ and $w_{z_\tau}^T (x_\tau)$ satisfies
\begin{align}
\bigg\langle \sum_{\tau=1}^t \Big( \nabla_{z_\tau} \ln w_{z_\tau}^T (x_\tau) \Big) \cdot \Delta z_\tau \bigg\rangle_{\mathrm{b}, 0 \to t} &\ge 0. \label{eq_condition_w_002_001}
\end{align}
where $\langle \cdot \rangle_{\mathrm{b}, \tau \to \tau'}$ is the expected value with respect to the backward path probability, that is, $\langle \cdot \rangle_{\mathrm{b}, \tau \to \tau'} \coloneqq \langle \cdot \rangle_{p_\mathrm{b} (X_{\tau \to \tau'}| \vec{Y}_{\tau + 1 \to \tau'}, \vec{Z}_{\tau + 1 \to \tau'})}$.
Then we have
\begin{align}
  \Phi_{0 \to t}^\mathrm{ex} &\le S (y_t, z_t) - S (y_0, z_0). \label{eq_Clausius_inequality_001_001}
\end{align}
Note that, when Eq.~\eqref{eq_condition_w_001_001} is satisfied, Eq.~\eqref{eq_condition_w_002_001} is also satisfied.

% !TEX root = ./paper_population_dynamics_000_001_main.tex

\section{Derivation of the first main result} \label{sec_proof_first_main_001}

This section is devoted to the derivation of {the Clausius equality, Eq.~\eqref{eq_Clausius_equality_001_001}}.
We first give another expression of {excess growth, Eq.~\eqref{eq_def_excess_growth_001_001},} that is, the {Berry-curvature-like} expression, and then the detailed derivation of {the Clausius equality, Eq.~\eqref{eq_Clausius_equality_001_001},} by using the {Berry-curvature-like} expression of {excess growth, Eq.~\eqref{eq_def_excess_growth_001_001}}.

\subsection{{Berry-curvature-like} expression of {excess growth}}

Before getting {the Clausius equality, Eq.~\eqref{eq_Clausius_equality_001_001},} we show the {Berry-curvature-like} expression of $\Phi_{0 \to t}^\mathrm{ex}$ in a quasi-static process:
\begin{align}
\Phi_{0 \to t}^\mathrm{ex} &= - \sum_{\tau=1}^t \bigg[ \sum_{x=1}^n u_{y_\tau, z_\tau}^\mathrm{st} (x) \nabla_{y_\tau} v_{y_\tau, z_\tau}^\mathrm{st} (x) \bigg] \cdot \Delta y_\tau - \sum_{\tau=1}^t \bigg[ \sum_{x=1}^n u_{y_\tau, z_\tau}^\mathrm{st} (x) \nabla_{z_\tau} v_{y_\tau, z_\tau}^\mathrm{st} (x) \bigg] \cdot \Delta z_\tau, \label{eq_excess_heat_discrete_Berry_curvature_001_001}
\end{align}
where $\nabla_y$ is the derivative with respect to $y$.
The proof of Eq.~\eqref{eq_excess_heat_discrete_Berry_curvature_001_001} is shown as follows.

\begin{proof}

By using the path integral formulation, {population growth, Eq.~\eqref{eq_population_growth_decomposition_001_001},} can be represented as
\begin{align}
e^{\Phi_{0 \to t}} &= \frac{1}{N_0} \sum_{\substack{\{ \vec{v}_{y_\tau, z_\tau}^\mathrm{eig} (x_\tau) \}_{x_\tau, \tau}, \\ \{ \vec{u}_{y_\tau, z_\tau}^\mathrm{eig} (x_\tau) \}_{x_\tau, \tau}}} \vec{1}_n \vec{v}_{y_t, z_t}^\mathrm{eig} (x_t) \bigg[ \prod_{\tau=1}^t \vec{u}_{y_\tau, z_\tau}^\mathrm{eig} (x_\tau) e^{\ln \hat{H}_{y_\tau, z_\tau}} \vec{v}_{y_{\tau-1}, z_{\tau-1}}^\mathrm{eig} (x_{\tau - 1}) \bigg] \vec{u}_{y_0, z_0}^\mathrm{eig} (x_0) \vec{N}_0, \label{eq_phi_all_001_001}
\end{align}
where $\vec{1}_n = [\underbrace{1, 1, \dots, 1}_{n}]$.
Note that $\vec{1}_n \vec{v}_{y_t, z_t}^\mathrm{eig} (x) = 1$ for $x = 1, 2, \dots, n$.

In a quasi-static process, {Eq.~\eqref{eq_phi_all_001_001}} is dominated by the eigenstate {that} has the largest eigenvalue; thus, when initial state is given by $\vec{v}_{y_0, z_0}^\mathrm{st}$, we have
\begin{align}
e^{\Phi_{0 \to t}} &= \frac{1}{N_0} \vec{1}_n \vec{v}_{y_t, z_t}^\mathrm{st} \bigg[ \prod_{\tau=1}^t \vec{u}_{y_\tau, z_\tau}^\mathrm{st} e^{\ln \hat{H}_{y_\tau, z_\tau}} \vec{v}_{y_{\tau-1}, z_{\tau-1}}^\mathrm{st} \bigg] \vec{u}_{y_0, z_0}^\mathrm{st} \vec{N}_0 \\
&= \prod_{\tau=1}^t \vec{u}_{y_\tau, z_\tau}^\mathrm{st} e^{\ln \hat{H}_{y_\tau, z_\tau}} \vec{v}_{y_{\tau-1}, z_{\tau-1}}^\mathrm{st}. \label{eq_phi_quasi-static_001_001}
\end{align}

Due to the fact that, for $x = 1, 2, \dots, n$,
{\begin{align}
u_{y, z}^\mathrm{st} (x) v_{y', z'}^\mathrm{st} (x) &= e^{- u_{y, z}^\mathrm{st} (x) \nabla_{y} v_{y, z}^\mathrm{st} (x) \cdot \Delta y} e^{- u_{y, z}^\mathrm{st} (x) \nabla_{z} v_{y, z}^\mathrm{st} (x) \cdot \Delta z},
\end{align}}%
each term in Eq.~\eqref{eq_phi_quasi-static_001_001} {is computed as}
{\begin{align}
\vec{u}_{y, z}^\mathrm{st} e^{\ln \hat{H}_{y, z}} \vec{v}_{y', z'}^\mathrm{st} &= e^{\phi_{y, z}^\mathrm{st}} e^{\sum_{x=1}^n u_{y, z}^\mathrm{st} (x) v_{y', z'}^\mathrm{st} (x)} \\
&= e^{\phi_{y, z}^\mathrm{st}} e^{- \sum_{x=1}^n u_{y, z}^\mathrm{st} (x) \nabla_{y} v_{y, z}^\mathrm{st} (x) \cdot \Delta y} e^{- \sum_{x=1}^n u_{y, z}^\mathrm{st} (x) \nabla_{z} v_{y, z}^\mathrm{st} (x) \cdot \Delta z}.
\end{align}}%
Then, Eq.~\eqref{eq_phi_quasi-static_001_001} can be rewritten as
\begin{align}
e^{\Phi_{0 \to t}} &= \prod_{\tau = 1}^t e^{\phi_{y_\tau, z_\tau}^\mathrm{st}} e^{- \sum_{x=1}^n u_{y_\tau, z_\tau}^\mathrm{st} (x) \nabla_{y_\tau} v_{y_\tau, z_\tau}^\mathrm{st} (x) \cdot \Delta y_\tau} e^{- \sum_{x=1}^n u_{y_\tau, z_\tau}^\mathrm{st} (x) \nabla_{z_\tau} v_{y_\tau, z_\tau}^\mathrm{st} (x) \cdot \Delta z_\tau}. \label{eq_phi_quasi-static_001_002}
\end{align}

By taking the logarithm of both sides of Eq.~\eqref{eq_phi_quasi-static_001_002}, we have
\begin{align}
\Phi_{0 \to t} &= \sum_{\tau = 1}^t \phi_{y_\tau, z_\tau}^\mathrm{st} - \sum_{\tau = 1}^t \sum_{x=1}^n u_{y_\tau, z_\tau}^\mathrm{st} (x) \nabla_{y_\tau} v_{y_\tau, z_\tau}^\mathrm{st} (x) \cdot \Delta y_\tau - \sum_{\tau = 1}^t \sum_{x=1}^n u_{y_\tau, z_\tau}^\mathrm{st} (x) \nabla_{z_\tau} v_{y_\tau, z_\tau}^\mathrm{st} (x) \cdot \Delta z_\tau.
\end{align}
Since $\Phi_{0 \to t}^\mathrm{hk} = \sum_{\tau = 1}^t \phi_{y_\tau, z_\tau}^\mathrm{st}$, we have $\Phi_{0 \to t}^\mathrm{ex}$, Eq.~\eqref{eq_excess_heat_discrete_Berry_curvature_001_001}.

\end{proof}

\subsection{Proof of the Clausius equality}

In this subsection, we provide the proof of the Clausius equality, Eq.~\eqref{eq_Clausius_equality_001_001}.
The proof is given as follows.

\begin{proof}

We define
\begin{align}
\tilde{H}_{y, z} (x| x') &\coloneqq \mu_y (x) T_z (x'| x).
\end{align}
Eq.~\eqref{eq_detailed_balance_condition_001_001} leads to
\begin{align}
\sum_{x'=1}^n \tilde{H}_{y, z} (x| x') \frac{v_{y, z}^\mathrm{st} (x')}{w_z^T (x')} &= \sum_{x'=1}^n \mu_y (x) T_z (x'| x) \frac{v_{y, z}^\mathrm{st} (x')}{w_z^T (x')} \\
&= \sum_{x'=1}^n \mu_y (x) T_z (x| x') \frac{v_{y, z}^\mathrm{st} (x')}{w_z^T (x)} \\
&= \sum_{x'=1}^n H_{y, z} (x| x') \frac{v_{y, z}^\mathrm{st} (x')}{w_z^T (x)} \\
&= e^{\phi_{y, z}^\mathrm{st}} \frac{v_{y, z}^\mathrm{st} (x)}{w_z^T (x)}.
\end{align}
Due to the assumption that {$T_z (x| x')$} is ergodic and {$H_{y, z} (x| x')$} is irreducible, and the Perron-Frobenius theorem, the largest left eigenvector is unique.
Thus, there exists $C_{y, z}$ such that, for $x = 1, 2, \dots, n$,
\begin{align}
u_{y, z}^\mathrm{st} (x) &= C_{y, z} \frac{v_{y, z}^\mathrm{st} (x)}{w_z^T (x)}. \label{eq_def_C_001_001}
\end{align}
By using the fact that $\| \vec{v}_{y, z}^\mathrm{st} \|_1 = 1$, we have
\begin{align}
C_{y, z} &= \sum_{x=1}^n w_z^T (x) u_{y, z}^\mathrm{st} (x).
\end{align}
Furthermore, since $\vec{u}_{y, z}^\mathrm{st} \vec{v}_{y, z}^\mathrm{st} = \sum_{x=1}^n u_{y, z}^\mathrm{st} (x) v_{y, z}^\mathrm{st} (x) = 1$, we obtain
\begin{align}
C_{y, z} &= \sum_{x=1}^n w_z^T (x) [ u_{y, z}^\mathrm{st} (x) ]^2. \label{eq_C_001_002}
\end{align}
By taking the derivative of Eq.~\eqref{eq_C_001_002} with respect to $y$, we have
\begin{align}
\frac{1}{2} \nabla_y \ln C_{y, z} &= \sum_{x=1}^n \frac{w_z^T (x) u_{y, z}^\mathrm{st} (x)}{C_{y, z}} \nabla_y u_{y, z}^\mathrm{st} (x) \\
&= \sum_{x=1}^n v_{y, z}^\mathrm{st} (x) \nabla_y u_{y, z}^\mathrm{st} (x) \\
&= - \sum_{x=1}^n u_{y, z}^\mathrm{st} (x) \nabla_y v_{y, z}^\mathrm{st} (x),
\end{align}
where we have used Eq.~\eqref{eq_def_C_001_001}.
Next, by taking the derivation of Eq.~\eqref{eq_C_001_002} with respect to $z$, we have almost the same expression:
\begin{align}
\frac{1}{2} \nabla_z \ln C_{y, z} &= \sum_{x=1}^n \frac{w_z^T (x) u_{y, z}^\mathrm{st} (x)}{C_{y, z}} \nabla_z u_{y, z}^\mathrm{st} (x) + \frac{1}{2} \sum_{x=1}^n \frac{(u_{y, z}^\mathrm{st} (x))^2}{C_{y, z}} \nabla_z w_z^T (x) \\
&= \sum_{x=1}^n v_{y, z}^\mathrm{st} (x) \nabla_z u_{y, z}^\mathrm{st} (x) + \frac{1}{2} \sum_{x=1}^n \frac{(u_{y, z}^\mathrm{st} (x))^2}{C_{y, z}} \nabla_z w_z^T (x) \\
&= \sum_{x=1}^n v_{y, z}^\mathrm{st} (x) \nabla_z u_{y, z}^\mathrm{st} (x) \label{eq_C_002_001} \\
&= - \sum_{x=1}^n u_{y, z}^\mathrm{st} (x) \nabla_z v_{y, z}^\mathrm{st} (x).
\end{align}
In Eq.~\eqref{eq_C_002_001}, the second term vanishes due to the assumption in Eq.~\eqref{eq_condition_w_001_001}.
Thus, Eq.~\eqref{eq_excess_heat_discrete_Berry_curvature_001_001} can be computed further as
\begin{align}
\Phi_{0 \to t}^\mathrm{ex} &= - \sum_{\tau=1}^t \bigg[ \sum_{x=1}^n u_{y_\tau, z_\tau}^\mathrm{st} (x) \nabla_{y_\tau} v_{y_\tau, z_\tau}^\mathrm{st} (x) \bigg] \cdot \Delta y_\tau - \sum_{\tau=1}^t \bigg[ \sum_{x=1}^n u_{y_\tau, z_\tau}^\mathrm{st} (x) \nabla_{z_\tau} v_{y_\tau, z_\tau}^\mathrm{st} (x) \bigg] \cdot \Delta z_\tau \\
&= \sum_{\tau=1}^t \bigg[ \frac{1}{2} \nabla_{y_\tau} \ln C_{y_\tau, z_\tau} \bigg] \cdot \Delta y_\tau + \sum_{\tau=1}^t \bigg[ \frac{1}{2} \nabla_{z_\tau} \ln C_{y_\tau, z_\tau} \bigg] \cdot \Delta z_\tau \\
&= \bigg[ \frac{1}{2} \ln C_{y, z} \bigg]_{y = y_0, z = z_0}^{y = y_t, z = z_t} \\
&= \big[ S_{y, z} \big]_{y = y_0, z = z_0}^{y = y_t, z = z_t} \\
&= S_{y_t, z_t} - S_{y_0, z_0},
\end{align}
where $S_{y, z} \coloneqq \frac{1}{2} \ln C_{y, z}$.
Therefore we have obtained the Clausius equality, Eq.~\eqref{eq_Clausius_equality_001_001}.

\end{proof}

% !TEX root = ./paper_population_dynamics_000_001_main.tex

\section{Derivation of the second main result} \label{sec_proof_second_main_001}

This section shows the derivation of {the Clausius inequality, Eq.~\eqref{eq_Clausius_inequality_001_001}}.
We first define two different path probabilities and a symmetrized divergence.
Then, we show the derivation of {the Clausius inequality, Eq.~\eqref{eq_Clausius_inequality_001_001}}.

\subsection{Path probabilities and symmetrized divergence}

In this subsection, we introduce the two different path probabilities and the symmetrized divergence on the path probabilities.
These quantities are essential to prove {the Clausius inequality, Eq.~\eqref{eq_Clausius_inequality_001_001}}.

\subsubsection{Retrospective process and its dual process}

Here, we introduce a Markov process that mimics the backward path probability for a fixed environment, called the retrospective process.
The backward path probability, Eq.~\eqref{eq_backward_path_probability_003_001}, is also expressed as
\begin{align}
p_\mathrm{b} (X_{0 \to t}| Y_{0 \to t}, Z_{0 \to t}) &= e^{- \sum_{\tau=1}^t \phi_\tau} \bigg[ \prod_{\tau=1}^t \vec{u} (x_\tau) e^{\ln \hat{H}_{y_\tau, z_\tau}} \vec{v} (x_{\tau-1}) \bigg] v_{y_0, z_0} (x_0) \\
&= \bigg[ \prod_{\tau=1}^t \vec{u} (x_\tau) e^{\ln \hat{H}_{y_\tau, z_\tau} - \phi_\tau \hat{1}_n} \vec{v} (x_{\tau-1}) \bigg] v_{y_0, z_0} (x_0).
\end{align}
Note that we have used $p_\mathrm{ini} (x_0) = v_{y_0, z_0} (x_0)$.
In a quasi-static process or the case where the changes of $y_\tau$ and $z_\tau$ are sufficiently small, we have
\begin{align}
p_\mathrm{b} (X_{0 \to t}| Y_{0 \to t}, Z_{0 \to t}) &\approx p_\mathrm{r} (X_{0 \to t}| Y_{0 \to t}, Z_{0 \to t}), \label{eq_relation_backward_retrospective_001_001}
\end{align}
where
\begin{align}
p_\mathrm{r} (X_{0 \to t}| Y_{0 \to t}, Z_{0 \to t}) &\coloneqq \bigg[ \prod_{\tau=1}^t \vec{u} (x_\tau) e^{\ln \hat{Q}_{y_\tau, z_\tau}} \vec{v} (x_{\tau-1}) \bigg] w_{y_0, z_0}^Q (x_0), \label{eq_retrospective_process_001_001}
\end{align}
and
\begin{align}
\hat{Q}_{y, z} &\coloneqq
\begin{bmatrix}
  Q_{y, z} (1| 1) & Q_{y, z} (1| 2) & \dots & Q_{y, z} (1| n) \\
  Q_{y, z} (2| 1) & Q_{y, z} (2| 2) & \dots & Q_{y, z} (2| n) \\
  \vdots & \vdots & \ddots & \vdots \\
  Q_{y, z} (n| 1) & Q_{y, z} (n| 2) & \dots & Q_{y, z} (3| n) \\
\end{bmatrix}, \label{eq_def_Q_001_001} \\
	Q_{y, z} (x| x') &\coloneqq \frac{u_{y, z}^\mathrm{st} (x)}{u_{y, z}^\mathrm{st} (x')} H_{y, z} (x| x') e^{- \phi_{y, z}^\mathrm{st}}. \label{eq_def_Q_001_002}
\end{align}
Note that
\begin{align}
\sum_{x=1}^n Q_{y, z} (x| x') &= \sum_{x=1}^n \frac{u_{y, z}^\mathrm{st} (x)}{u_{y, z}^\mathrm{st} (x')} H_{y, z} (x| x') e^{- \phi_{y, z}^\mathrm{st}} \\
&= \frac{u_{y, z}^\mathrm{st} (x')}{u_{y, z}^\mathrm{st} (x')} e^{\phi_{y, z}^\mathrm{st}} e^{- \phi_{y, z}^\mathrm{st}} \\
&= 1; \label{eq_Q_transition_matrix_001}
\end{align}
thus, {$Q_{y, z} (x| x')$} can be interpreted as a conditional probability.

We then consider the Markov process induced by
\begin{align}
\tilde{Q}_{y, z} (x| x') &\coloneqq \frac{v_{y, z}^\mathrm{st} (x)}{v_{y, z}^\mathrm{st} (x')} H_{y, z} (x'| x) e^{- \phi_{y, z}^\mathrm{st}}. \label{eq_def_Q_tilde_001_001}
\end{align}
Note that the following identity holds:
\begin{align}
\sum_{x=1}^n \tilde{Q}_{y, z} (x| x') &= \sum_{x=1}^n \frac{v_{y, z}^\mathrm{st} (x)}{v_{y, z}^\mathrm{st} (x')} H_{y, z} (x'| x) e^{- \phi_{y, z}^\mathrm{st}} \\
&= 1.
\end{align}
From Eqs.~\eqref{eq_def_Q_001_002} and \eqref{eq_def_Q_tilde_001_001}, we have
\begin{align}
\tilde{Q}_{y, z} (x| x') &= \frac{v_{y, z}^\mathrm{st} (x)}{v_{y, z}^\mathrm{st} (x')} H_{y, z} (x'| x) e^{- \phi_{y, z}^\mathrm{st}} \\
&= \frac{w_{y, z}^Q (x)}{w_{y, z}^Q (x')} \bigg( \frac{u_{y, z}^\mathrm{st} (x')}{u_{y, z}^\mathrm{st} (x)} H_{y, z} (x'| x) e^{- \phi_{y, z}^\mathrm{st}} \bigg) \\
&= \frac{w_{y, z}^Q (x)}{w_{y, z}^Q (x')} Q_{y, z} (x'| x),
\end{align}
where
\begin{align}
w_{y, z}^Q (x) &\coloneqq v_{y, z}^\mathrm{st} (x) u_{y, z}^\mathrm{st} (x). \label{eq_def_wQ_001_001}
\end{align}
We also note that $w_{y, z}^Q (x)$ satisfies the condition of probability distribution: $\sum_{x=1}^n w_{y, z}^Q (x) = 1$.

The process induced by Eq.~\eqref{eq_def_Q_tilde_001_001}, which we call the dual retrospective process, is given by
\begin{align}
p_\mathrm{dr} (X_{0 \to t}^\mathrm{r}| Y_{0 \to t}^\mathrm{r}, Z_{0 \to t}^\mathrm{r}) &\coloneqq {\bigg[ \prod_{\tau=1}^t \vec{u} (x_{\tau - 1}) e^{\ln \hat{\tilde{Q}}_{y_\tau, z_\tau}} \vec{v} (x_\tau) \bigg] w_{y_t, z_t}^Q (x_t),} \label{eq_dual_retrospective_process_001_001}
\end{align}
where
\begin{align}
  \hat{\tilde{Q}}_{y, z} &\coloneqq
  \begin{bmatrix}
    \tilde{Q}_{y, z} (1| 1) & \tilde{Q}_{y, z} (1| 2) & \dots & \tilde{Q}_{y, z} (1| n) \\
    \tilde{Q}_{y, z} (2| 1) & \tilde{Q}_{y, z} (2| 2) & \dots & \tilde{Q}_{y, z} (2| n) \\
    \vdots & \vdots & \ddots & \vdots \\
    \tilde{Q}_{y, z} (n| 1) & \tilde{Q}_{y, z} (n| 2) & \dots & \tilde{Q}_{y, z} (n| n) \\
  \end{bmatrix}, \label{eq_def_Q_tilde_002_001}
\end{align}
and $X_{0 \to t}^\mathrm{r}$, $Y_{0 \to t}^\mathrm{r}$, and $Z_{0 \to t}^\mathrm{r}$ are the reversed processes of $X_{0 \to t}$, $Y_{0 \to t}$, and $Z_{0 \to t}$, respectively.

\subsubsection{Symmetrized divergence}

We define the symmetrized divergence on the path probabilities by computing the logarithm of the ratios between path probabilities.
In the following subsection, we utilize it to prove the Clausius inequality, Eq.~\eqref{eq_Clausius_inequality_001_001}.

From Eq.~\eqref{eq_def_Q_001_001}, we have
{\begin{align}
\vec{u} (x) e^{\ln \hat{Q}_{y, z}} \vec{v} (x') &= e^{- \phi_{y, z}^\mathrm{st}} \frac{u_{y, z}^\mathrm{st} (x)}{u_{y, z}^\mathrm{st} (x_{\tau - 1})} \vec{u} (x) e^{\ln \hat{H}_{y, z}} \vec{v} (x').
\end{align}}%
Then, Eq.~\eqref{eq_retrospective_process_001_001} can be rewritten as
\begin{align}
p_\mathrm{r} (X_{0 \to t}| Y_{0 \to t}, Z_{0 \to t}) &= \bigg[ \prod_{\tau=1}^t \vec{u} (x_\tau) e^{\ln \hat{Q}_{y_\tau, z_\tau}} \vec{v} (x_{\tau-1}) \bigg] v_{y_0, z_0} (x_0) \\
&= e^{- \sum_\tau \phi_{y_\tau, z_\tau}^\mathrm{st}} \frac{\prod_{\tau=1}^t u_{y_\tau, z_\tau}^\mathrm{st} (x_\tau)}{\prod_{\tau=1}^t u_{y_\tau, z_\tau}^\mathrm{st} (x_{\tau - 1})} \bigg[ \prod_{\tau=1}^t \vec{u} (x_\tau) e^{\ln \hat{H}_{y_\tau, z_\tau}} \vec{v} (x_{\tau-1}) \bigg] w_{y_0, z_0}^Q (x_0),
\end{align}
and
\begin{align}
p_\mathrm{r} (X_{0 \to t}| Y_{0 \to t}, Z_{0 \to t}) &= \exp \bigg( - \sum_{\tau=1}^t \phi_{y_\tau, z_\tau}^\mathrm{st} \bigg) \exp \bigg( - \sum_{\tau=1}^t \big[ \nabla_{y_\tau} \ln u_{y_\tau, z_\tau}^\mathrm{st} (x_\tau) \big] \cdot \Delta y_\tau \bigg) \exp \bigg( - \sum_{\tau=1}^t \big[ \nabla_{z_\tau} \ln u_{y_\tau, z_\tau}^\mathrm{st} (x_\tau) \big] \cdot \Delta z_\tau \bigg) \nonumber \\
& \quad \times \exp \bigg( \ln \frac{u_{y_t, z_t}^\mathrm{st} (x_t)}{u_{y_0, z_0}^\mathrm{st} (x_0)} + \ln \frac{w_{y_0, z_0}^Q (x_0)}{v_{y_0, z_0}^\mathrm{st} (x_0)} \bigg) e^{\Phi_{0 \to t}} p_\mathrm{b} (X_{0 \to t}| Y_{0 \to t}, Z_{0 \to t}).
\end{align}
Thus, the logarithm of the ratio between $p_\mathrm{b} (X_{0 \to t}| Y_{0 \to t}, Z_{0 \to t})$ and $p_\mathrm{r} (X_{0 \to t}| Y_{0 \to t}, Z_{0 \to t})$ is computed as
\begin{align}
\ln \frac{p_\mathrm{b} (X_{0 \to t}| Y_{0 \to t}, Z_{0 \to t})}{p_\mathrm{r} (X_{0 \to t}| Y_{0 \to t}, Z_{0 \to t})} &= - \Phi_{0 \to t}^\mathrm{ex} - \ln \frac{u_{y_t, z_t}^\mathrm{st} (x_t)}{u_{y_0, z_0}^\mathrm{st} (x_0)} - \ln \frac{w_{y_0, z_0}^Q (x_0)}{v_{y_0, z_0}^\mathrm{st} (x_0)} \nonumber \\
& \quad + \sum_{\tau=1}^t \big[ \nabla_{y_\tau} \ln u_{y_\tau, z_\tau}^\mathrm{st} (x_\tau) \big] \cdot \Delta y_\tau + \sum_{\tau=1}^t \big[ \nabla_{z_\tau} \ln u_{y_\tau, z_\tau}^\mathrm{st} (x_\tau) \big] \cdot \Delta z_\tau \\
&= - \Phi_{0 \to t}^\mathrm{ex} - \ln u_{y_t, z_t}^\mathrm{st} (x_t) + \sum_{\tau=1}^t \big[ \nabla_{y_\tau} \ln u_{y_\tau, z_\tau}^\mathrm{st} (x_\tau) \big] \cdot \Delta y_\tau + \sum_{\tau=1}^t \big[ \nabla_{z_\tau} \ln u_{y_\tau, z_\tau}^\mathrm{st} (x_\tau) \big] \cdot \Delta z_\tau, \label{eq_log_ratio_001_001}
\end{align}
where we have used Eqs.~\eqref{eq_def_housekeeping_growth_001_001} and \eqref{eq_def_excess_growth_001_001}.

Next, we consider the dual retrospective process.
From Eq.~\eqref{eq_def_Q_tilde_002_001}, we have
{\begin{align}
\vec{u} (x') e^{\ln \hat{\tilde{Q}}_{y, z}} \vec{v} (x) &= e^{- \phi_{y, z}^\mathrm{st}} \frac{v_{y, z}^\mathrm{st} (x_{\tau - 1})}{v_{y, z}^\mathrm{st} (x)} \vec{u} (x) e^{\ln \hat{H}_{y, z}} \vec{v} (x').
\end{align}}%
Then, Eq.~\eqref{eq_dual_retrospective_process_001_001} is rewritten as
\begin{align}
p_\mathrm{dr} (X_{0 \to t}^\mathrm{r}| Y_{0 \to t}^\mathrm{r}, Z_{0 \to t}^\mathrm{r}) &= e^{- \sum_\tau \phi_{y_\tau, z_\tau}^\mathrm{st}} \frac{\prod_{\tau=1}^t v_{y_\tau, z_\tau}^\mathrm{st} (x_{\tau-1})}{\prod_{\tau=1}^t v_{y_\tau, z_\tau}^\mathrm{st} (x_\tau)} \bigg[ \prod_{\tau=1}^t \vec{u} (x_\tau) e^{\ln \hat{H}_{y_\tau, z_\tau}} \vec{v} (x_{\tau-1}) \bigg] w_{y_t, z_t}^Q (x_t),
\end{align}
and {then we have}
\begin{align}
p_\mathrm{dr} (X_{0 \to t}^\mathrm{r}| Y_{0 \to t}^\mathrm{r}, Z_{0 \to t}^\mathrm{r}) &= \exp \bigg( - \sum_{\tau=1}^t \phi_{y_\tau, z_\tau}^\mathrm{st} \bigg) \exp \bigg( \sum_{\tau=1}^t \big[ \nabla_{y_\tau} \ln v_{y_\tau, z_\tau}^\mathrm{st} (x_\tau) \big] \cdot \Delta y_\tau \bigg) \exp \bigg( \sum_{\tau=1}^t \big[ \nabla_{z_\tau} \ln v_{y_\tau, z_\tau}^\mathrm{st} (x_\tau) \big] \cdot \Delta z_\tau \bigg) \nonumber \\
& \quad \times \exp \bigg( \ln \frac{v_{y_0, z_0}^\mathrm{st} (x_0)}{v_{y_t, z_t}^\mathrm{st} (x_t)} + \ln \frac{w_{y_t, z_t}^Q (x_t)}{v_{y_0, z_0}^\mathrm{st} (x_0)} \bigg) e^{\Phi_{0 \to t}} p_\mathrm{b} (X_{0 \to t}| Y_{0 \to t}, Z_{0 \to t}).
\end{align}
Thus, the logarithm of the ratio between $p_\mathrm{b} (X_{0 \to t}| Y_{0 \to t}, Z_{0 \to t})$ and $p_\mathrm{dr} (X_{0 \to t}^\mathrm{r}| Y_{0 \to t}^\mathrm{r}, Z_{0 \to t}^\mathrm{r})$ is computed as
\begin{align}
\ln \frac{p_\mathrm{b} (X_{0 \to t}| Y_{0 \to t}, Z_{0 \to t})}{p_\mathrm{dr} (X_{0 \to t}^\mathrm{r}| Y_{0 \to t}^\mathrm{r}, Z_{0 \to t}^\mathrm{r})} &= - \Phi_{0 \to t}^\mathrm{ex} - \ln \frac{v_{y_0, z_0}^\mathrm{st} (x_0)}{v_{y_t, z_t}^\mathrm{st} (x_t)} - \ln \frac{w_{y_t, z_t}^Q (x_t)}{v_{y_0, z_0}^\mathrm{st} (x_0)} \nonumber \\
& \quad - \sum_{\tau=1}^t \big[ \nabla_{y_\tau} \ln v_{y_\tau, z_\tau}^\mathrm{st} (x_\tau) \big] \cdot \Delta y_\tau - \sum_{\tau=1}^t \big[ \nabla_{z_\tau} \ln v_{y_\tau, z_\tau}^\mathrm{st} (x_\tau) \big] \cdot \Delta z_\tau \\
&= - \Phi_{0 \to t}^\mathrm{ex} - \ln u_{y_t, z_t}^\mathrm{st} (x_t) - \sum_{\tau=1}^t \big[ \nabla_{y_\tau} \ln v_{y_\tau, z_\tau}^\mathrm{st} (x_\tau) \big] \cdot \Delta y_\tau - \sum_{\tau=1}^t \big[ \nabla_{z_\tau} \ln v_{y_\tau, z_\tau}^\mathrm{st} (x_\tau) \big] \cdot \Delta z_\tau. \label{eq_log_ratio_001_002}
\end{align}

By using Eqs.~\eqref{eq_log_ratio_001_001} and \eqref{eq_log_ratio_001_002}, we have
\begin{align}
& \frac{1}{2} \ln \frac{p_\mathrm{b} (X_{0 \to t}| Y_{0 \to t}, Z_{0 \to t})}{p_\mathrm{r} (X_{0 \to t}| Y_{0 \to t}, Z_{0 \to t})} + \frac{1}{2} \ln \frac{p_\mathrm{b} (X_{0 \to t}| Y_{0 \to t}, Z_{0 \to t})}{p_\mathrm{dr} (X_{0 \to t}^\mathrm{r}| Y_{0 \to t}^\mathrm{r}, Z_{0 \to t}^\mathrm{r})} \nonumber \\
& \quad = \ln \frac{p_\mathrm{b} (X_{0 \to t}| Y_{0 \to t}, Z_{0 \to t})}{\big[ p_\mathrm{r} (X_{0 \to t}| Y_{0 \to t}, Z_{0 \to t}) p_\mathrm{dr} (X_{0 \to t}^\mathrm{r}| Y_{0 \to t}^\mathrm{r}, Z_{0 \to t}^\mathrm{r}) \big]^{\frac{1}{2}}} \\
& \quad = - \Phi_{0 \to t}^\mathrm{ex} - \ln u_{y_t, z_t}^\mathrm{st} (x_t) + \frac{1}{2} \sum_{\tau=1}^t \bigg[ \nabla_{y_\tau} \ln \frac{u_{y_\tau, z_\tau}^\mathrm{st} (x_\tau)}{v_{y_\tau, z_\tau}^\mathrm{st} (x_\tau)} \bigg] \cdot \Delta y_\tau + \frac{1}{2} \sum_{\tau=1}^t \bigg[ \nabla_{z_\tau} \ln \frac{u_{y_\tau, z_\tau}^\mathrm{st} (x_\tau)}{v_{y_\tau, z_\tau}^\mathrm{st} (x_\tau)} \bigg] \cdot \Delta z_\tau \\
& \quad = - \Phi_{0 \to t}^\mathrm{ex} - \ln u_{y_t, z_t}^\mathrm{st} (x_t) + \frac{1}{2} \sum_{\tau=1}^t \bigg[ \nabla_{y_\tau} \ln \frac{C_{y_\tau, z_\tau}}{w_{z_\tau}^T (x_\tau)} \bigg] \cdot \Delta y_\tau + \frac{1}{2} \sum_{\tau=1}^t \bigg[ \nabla_{z_\tau} \ln \frac{C_{y_\tau, z_\tau}}{w_{z_\tau}^T (x_\tau)} \bigg] \cdot \Delta z_\tau \\
& \quad = - \Phi_{0 \to t}^\mathrm{ex} - \ln u_{y_t, z_t}^\mathrm{st} (x_t) + \frac{1}{2} \big[ \ln C_{y_t, z_t} - \ln C_{y_0, z_0} \big] - \frac{1}{2} \sum_{\tau=1}^t \Big( \nabla_{z_\tau} \ln w_{z_\tau}^T (x_\tau) \Big) \cdot \Delta z_\tau.
\end{align}

Then, we define the symmetrized divergence as
\begin{align}
D_{0 \to t}^\mathrm{sym} &\coloneqq \bigg\langle \ln \frac{p_\mathrm{b} (X_{0 \to t}| Y_{0 \to t}, Z_{0 \to t})}{\big[ p_\mathrm{r} (X_{0 \to t}| Y_{0 \to t}, Z_{0 \to t}) p_\mathrm{dr} (X_{0 \to t}^\mathrm{r}| Y_{0 \to t}^\mathrm{r}, Z_{0 \to t}^\mathrm{r}) \big]^{\frac{1}{2}}} \bigg\rangle_{\mathrm{b}, 0 \to t}, \label{eq_def_divergence_sym_001_001}
\end{align}
and it is computed as
\begin{align}
D_{0 \to t}^\mathrm{sym} &= \int \mathcal{D} X_{0 \to t} \, p_\mathrm{b} (X_{0 \to t}| Y_{0 \to t}, Z_{0 \to t}) \ln \frac{p_\mathrm{b} (X_{0 \to t}| Y_{0 \to t}, Z_{0 \to t})}{\big[ p_\mathrm{r} (X_{0 \to t}| Y_{0 \to t}, Z_{0 \to t}) p_\mathrm{dr} (X_{0 \to t}^\mathrm{r}| Y_{0 \to t}^\mathrm{r}, Z_{0 \to t}^\mathrm{r}) \big]^{\frac{1}{2}}} \\
&= - \Phi_{0 \to t}^\mathrm{ex} + S_{y_t, z_t} - S_{y_0, z_0} - \int \mathcal{D} X_{0 \to t} \, p_\mathrm{b} (X_{0 \to t}| Y_{0 \to t}, Z_{0 \to t}) \ln u_{y_t, z_t}^\mathrm{st} (x_t) \\
&= - \Phi_{0 \to t}^\mathrm{ex} + S_{y_t, z_t} - S_{y_0, z_0} - \sum_{x_t=1}^n v_{y_t, z_t}^\mathrm{st} (x_t) \ln u_{y_t, z_t}^\mathrm{st} (x_t) - \bigg\langle \sum_{\tau=1}^t \Big( \nabla_{z_\tau} \ln w_{z_\tau}^T (x_\tau) \Big) \cdot \Delta z_\tau \bigg\rangle_{\mathrm{b}, 0 \to t}, \label{eq_symmetrized_divergence_001_001}
\end{align}
where $\int \mathcal{D} X_{0 \to t} \coloneqq \sum_{\{ X_{0 \to t}\}}$.
Note that we have used $v_{y_t, z_t}^\mathrm{st} (x_t) = \sum_{x_0, x_1, \dots, x_{t-1}} p_\mathrm{b} (X_{0 \to t}| Y_{0 \to t}, Z_{0 \to t})$ and $v_{y_0, z_0}^\mathrm{st} (x_0) = \sum_{x_1, x_2, \dots, x_t} p_\mathrm{b} (X_{0 \to t}| Y_{0 \to t}, Z_{0 \to t})$ since we are studying a transition between stationary states. % that is, at time $t$, the system is in a stationary state.
Note that the non-negativity of the symmetrized divergence is shown in Appendix~\ref{app_sec_nonnegativity_divergence_001_001}.

\subsection{Proof of the Clausius inequality}

We define
\begin{align}
  \sigma_{0 \to t} &\coloneqq - \Phi_{0 \to t}^\mathrm{ex} + S_{y_t, z_t} - S_{y_0, z_0}. \label{eq_def_entropy_production_001_001}
\end{align}
Note that $\sigma_{0 \to t}$ is not thermodynamically the entropy production, but we call it the entropy production since its mathematical properties are similar to the thermodynamic entropy production.
In the rest of this section, we prove the following inequality:
\begin{align}
  \sigma_{0 \to t} &\ge 0. \label{eq_entropy_production_001_001}
\end{align}
Eq.~\eqref{eq_entropy_production_001_001} leads to {the Clausius inequality, Eq.~\eqref{eq_Clausius_inequality_001_001}}.

\begin{proof}

{For any given process from $\tau=0$ to $\tau=t$, we can define a new process from $\tau = 0$ to $\tau = t'$ such that it is identical to the given process from $\tau = 0$ to $\tau = t$ and the duplication rate of the new process is constant ($\mu_{y_{\tau}} (x) = \mathrm{const.}$) for $t \le \tau \le t'$.
In this proof, we use the new process.
Note that we can perform this technique for any process; so, this discussion is not a limitation.}
To describe the new process, we also define $Y_{0 \to t'} \coloneqq \{ y_\tau \}_{\tau=0}^{t'}$, $Y_{t + 1 \to t'} \coloneqq \{ y_\tau \}_{\tau=t+1}^{t'}$, $Z_{0 \to t'} \coloneqq \{ z_\tau \}_{\tau = 0}^{t'}$, and $Z_{t \to t'} \coloneqq \{ z_\tau \}_{\tau=t+1}^{t'}$.

From Eq.~\eqref{eq_symmetrized_divergence_001_001}, we have
\begin{align}
\Phi_{0 \to t}^\mathrm{ex} &= S_{y_t, z_t} - S_{y_0, z_0} - D_{0 \to t}^\mathrm{sym} - \sum_{x_t=1}^n v_{y_t, z_t}^\mathrm{st} (x_t) \ln u_{y_t, z_t}^\mathrm{st} (x_t) - \bigg\langle \sum_{\tau=1}^t \Big( \nabla_{z_\tau} \ln w_{z_\tau}^T (x_\tau) \Big) \cdot \Delta z_\tau \bigg\rangle_{\mathrm{b}, 0 \to t}.
\end{align}
{Due to the assumption $\mu_{y_{t'}} (x) = \mathrm{const.}$, we have $u_{y_{t'}, z_{t'}}^\mathrm{st} (x) = 1$ for $x = 1, 2, \dots, n$.
Then, we get $S_{y_{t'}, z_{t'}} = 0$ and $\sum_{x_{t'}=1}^n v_{y_{t'}, z_{t'}}^\mathrm{st} (x_{t'}) \ln u_{y_{t'}, z_{t'}}^\mathrm{st} (x_{t'}) = 0$~\cite{Sughiyama_002};} thus we obtain
\begin{align}
\Phi_{0 \to t'}^\mathrm{ex} &= S_{y_{t'}, z_{t'}} - S_{y_0, z_0} - D_{0 \to t'}^{\mathrm{sym}} - \sum_{i_{t'}=1}^n v_{y_{t'}, z_{t'}}^\mathrm{st} (x_{t'}) \ln u_{y_{t'}, z_{t'}}^\mathrm{st} (x_{t'}) - \bigg\langle \sum_{\tau=1}^{t'} \Big( \nabla_{z_\tau} \ln w_{z_\tau}^T (x_\tau) \Big) \cdot \Delta z_\tau \bigg\rangle_{\mathrm{b}, 0 \to t'} \\
&= - S_{y_0, z_0} - D_{0 \to t'}^{\mathrm{sym}} - \bigg\langle \sum_{\tau=1}^{t'} \Big( \nabla_{z_\tau} \ln w_{z_\tau}^T (x_\tau) \Big) \cdot \Delta z_\tau \bigg\rangle_{\mathrm{b}, 0 \to t'} \\
&\le - S_{y_0, z_0} - \bigg\langle \sum_{\tau=1}^{t'} \Big( \nabla_{z_\tau} \ln w_{z_\tau}^T (x_\tau) \Big) \cdot \Delta z_\tau \bigg\rangle_{\mathrm{b}, 0 \to t'},
\end{align}
and
\begin{align}
\Phi_{t \to t'}^\mathrm{ex} &= S_{y_{t'}, z_{t'}} - S_{y_t, z_t} - D_{t \to t'}^\mathrm{sym} - \sum_{i_{t'}=1}^n v_{y_{t'}, z_{t'}}^\mathrm{st} (x_{t'}) \ln u_{y_{t'}, z_{t'}}^\mathrm{st} (x_{t'}) - \bigg\langle \sum_{\tau=t+1}^{t'} \Big( \nabla_{z_\tau} \ln w_{z_\tau}^T (x_\tau) \Big) \cdot \Delta z_\tau \bigg\rangle_{\mathrm{b}, t \to t'} \\
&= - S_{y_t, z_t} - D_{t \to t'}^\mathrm{sym} - \bigg\langle \sum_{\tau=t+1}^{t'} \Big( \nabla_{z_\tau} \ln w_{z_\tau}^T (x_\tau) \Big) \cdot \Delta z_\tau \bigg\rangle_{\mathrm{b}, t \to t'} \\
&\le - S_{y_t, z_t} - \bigg\langle \sum_{\tau=t+1}^{t'} \Big( \nabla_{z_\tau} \ln w_{z_\tau}^T (x_\tau) \Big) \cdot \Delta z_\tau \bigg\rangle_{\mathrm{b}, t \to t'}.
\end{align}

Next, let us denote $Y_{0 \to t}$, $Y_{t + 1 \to t'}$, and $Y_{1 \to t'}$ that maximize $\Phi_{0 \to t}^\mathrm{ex}$, $\Phi_{t \to t'}^\mathrm{ex}$, and $\Phi_{0 \to t'}^\mathrm{ex}$, by $Y_{0 \to t}^*$, $Y_{t + 1 \to t'}^*$, and $Y_{1 \to t'}^*$, respectively.
We also denote the maximized values of $\Phi_{0 \to t}^\mathrm{ex}$, $\Phi_{t \to t'}^\mathrm{ex}$, $\Phi_{0 \to t'}^\mathrm{ex}$, and $D_{0 \to t}^\mathrm{sym}$ by $\Phi_{0 \to t}^{\mathrm{ex}, *}$, $\Phi_{t \to t'}^{\mathrm{ex}, *}$, $\Phi_{0 \to t'}^{\mathrm{ex}, *}$, and $D_{0 \to t}^{\mathrm{sym}, *}$, respectively.
In this case, the following relation holds:
\begin{align}
\Phi_{0 \to t}^{\mathrm{ex}, *} + \Phi_{t \to t'}^{\mathrm{ex}, *} &\le \Phi_{0 \to t'}^{\mathrm{ex}, *}, \label{eq_inequality_excess_heat_001_001}
\end{align}
and, due to the Clausius equality, Eq.~\eqref{eq_Clausius_equality_001_001}, we have
\begin{align}
\Phi_{t \to t'}^{\mathrm{ex}, *} &= - S_{y_t, z_t} - \bigg\langle \sum_{\tau=t+1}^{t'} \Big( \nabla_{z_\tau} \ln w_{z_\tau}^T (x_\tau) \Big) \cdot \Delta z_\tau \bigg\rangle_{\mathrm{b}, t \to t'}, \\
\Phi_{0 \to t'}^{\mathrm{ex}, *} &= - S_{y_0, z_0} - \bigg\langle \sum_{\tau=1}^{t'} \Big( \nabla_{z_\tau} \ln w_{z_\tau}^T (x_\tau) \Big) \cdot \Delta z_\tau \bigg\rangle_{\mathrm{b}, 0 \to t'}.
\end{align}
Thus, Eq.~\eqref{eq_inequality_excess_heat_001_001} leads to
\begin{align}
S_{y_t, z_t} - S_{y_0, z_0} - D_{0 \to t}^{\mathrm{sym}, *} - \sum_{x_t=1}^n v_{y_t, z_t}^\mathrm{st} (x_t) \ln u_{y_t, z_t}^\mathrm{st} (x_t) - S_{y_t, z_t} &\le - S_{y_0, z_0}.
\end{align}
Then,
\begin{align}
  D_{0 \to t}^{\mathrm{sym}, *} + \sum_{x_t=1}^n v_{y_t, z_t}^\mathrm{st} (x_t) \ln u_{y_t, z_t}^\mathrm{st} (x_t) &\ge 0. \label{eq_proof_Clausius_inequality_001_001}
\end{align}
By definition, $Y_{0 \to t}^*$ maximizes $\Phi_{0 \to t}^\mathrm{ex}$; thus, for any $Y_{0 \to t}$, we have
\begin{align}
  D_{0 \to t}^\mathrm{sym} &\ge D_{0 \to t}^{\mathrm{sym}, *}. \label{eq_proof_Clausius_inequality_002_001}
\end{align}

By using Eqs.~\eqref{eq_def_entropy_production_001_001} and \eqref{eq_symmetrized_divergence_001_001}, we have
\begin{align}
\sigma_{0 \to t} &= D_{0 \to t}^\mathrm{sym} + \sum_{x_t=1}^n v_{y_t, z_t}^\mathrm{st} (x_t) \ln u_{y_t, z_t}^\mathrm{st} (x_t)  + \bigg\langle \sum_{\tau=1}^t \Big( \nabla_{z_\tau} \ln w_{z_\tau}^T (x_\tau) \Big) \cdot \Delta z_\tau \bigg\rangle_{\mathrm{b}, 0 \to t}.
\end{align}
Thus, Eqs.~\eqref{eq_proof_Clausius_inequality_001_001}, \eqref{eq_proof_Clausius_inequality_002_001}, and \eqref{eq_condition_w_002_001} lead to Eq.~\eqref{eq_entropy_production_001_001}.

\end{proof}

% !TEX root = ./paper_population_dynamics_000_001_main.tex

\section{Numerical simulations} \label{sec_numerical_simulation_001}

In this section, we show numerical simulations to demonstrate the second main result, Eq.~\eqref{eq_Clausius_inequality_001_001}.
Using Eqs.~\eqref{eq_def_excess_growth_001_001}, we transform \eqref{eq_Clausius_inequality_001_001} into
\begin{align}
  \Phi_{0 \to t} &\le \Phi_{0 \to t}^\mathrm{hk} + S_{y_t, z_t} - S_{y_0, z_0}. \label{eq_inequality_checked_001_001}
\end{align}
Furthermore, Eq.~\eqref{eq_inequality_checked_001_002} is simplified as follows when $t$ is sufficiently long or $y_0 = y_t$ and $z_0 = z_t$ are satisfied:
\begin{align}
  \Phi_{0 \to t} &\le \Phi_{0 \to t}^\mathrm{hk}. \label{eq_inequality_checked_001_002}
\end{align}
Then we confirm Eq.~\eqref{eq_inequality_checked_001_002} {via numerical simulations}.
Note that, in the setup, $t$ is sufficiently large.

\subsection{Problem setup of numerical simulations}

For simplicity, we consider organisms of two phenotypes.
We set {the duplication rates} of two phenotypes $\mu_{y_\tau} (1) = 0$ and $\mu_{y_\tau} (2) = y_\tau$, where $\Delta y_\tau \coloneqq y_\tau - y_{\tau - 1} \sim \mathcal{N} (\cdot; m_y, (\sigma_y)^2)$ for $\tau = 1, 2, \dots, t$.
We consider the following transition matrices:
\begin{subequations} \label{eq_transition_example_001_001}
\begin{align}
  \hat{T}_{z_\tau > 0} &=
  \begin{bmatrix}
    1 - \omega_\alpha & \omega_\alpha \\
    \omega_\alpha & 1 - \omega_\alpha
  \end{bmatrix}, \\
  \hat{T}_{z_\tau < 0} &=
  \begin{bmatrix}
    1 - \omega_\beta & \omega_\beta \\
    \omega_\beta & 1 - \omega_\beta
  \end{bmatrix}, \\
  \hat{T}_{z_\tau = 0} &=
  \begin{bmatrix}
    1 - \omega_\gamma & \omega_\gamma \\
    \omega_\gamma & 1 - \omega_\gamma
  \end{bmatrix},
\end{align} \label{eq_transition_matrix_numerical_simulation_001_001}%
\end{subequations}
where $z_\tau \sim \mathcal{N} (\cdot; y_\tau, (\sigma_z)^2)$ and $1 > \omega_\alpha, \omega_\beta \ge 0$.
We also put $\omega_\alpha \ge \omega_\beta$ and $\omega_\gamma \coloneqq \frac{1}{2} (\omega_\alpha + \omega_\beta)$.
Note that Eq.~\eqref{eq_transition_example_001_001} has the same form as Eq.~\eqref{eq_example_T_001_001}; thus, Eq.~\eqref{eq_transition_matrix_numerical_simulation_001_001} satisfies the conditions of the Clausius equality and inequality, Eqs.~\eqref{eq_condition_w_001_001} and \eqref{eq_condition_w_002_001}.

So far, we have discussed using the temporal information $z_\tau$ at time $\tau$ and investigated this in Sec.~\ref{sec_numerical_simulation_001_001}.
We refer to this as the naive strategy.
We also consider another approach.
In Sec.~\ref{sec_numerical_simulation_001_001}, we deal with the time-averaging of $z_\tau$ given by
\begin{align}
\zeta_\tau (t^\mathrm{ta}) &\coloneqq \frac{1}{t^\mathrm{ta}} \sum_{\tau' = \tau - t^\mathrm{ta} + 1}^\tau z_{\tau'}, \label{eq_moving_average_001_001}
\end{align}
and $z_\tau = 0$ for $\tau \le 0$; that is, $\hat{T}_{\zeta_\tau (t^\mathrm{ta})}$ is used instead of $\hat{T}_{z_\tau}$.
We refer to this as the time-averaging strategy.

\subsection{Numerical simulation I} \label{sec_numerical_simulation_001_001}

Here we focus on the naive strategy.
We set $m_y = 0.0$ and $\sigma_y = 0.0010$ for the distribution of $\Delta y_\tau$, and use $\omega_\alpha = 0.05$ and $\omega_\beta = 0.01$ for $T_{z_\tau} (x| x')$.
We start with $N_0 (1) = N_0 (2) = 5.0$.
To see the effect of noise on information, we consider two distributions {of} $z_\tau$: $\sigma_z = 0.010$ and $\sigma_z = 0.0010$.

\begin{figure}[tb]
\centering
\includegraphics[scale=0.50]{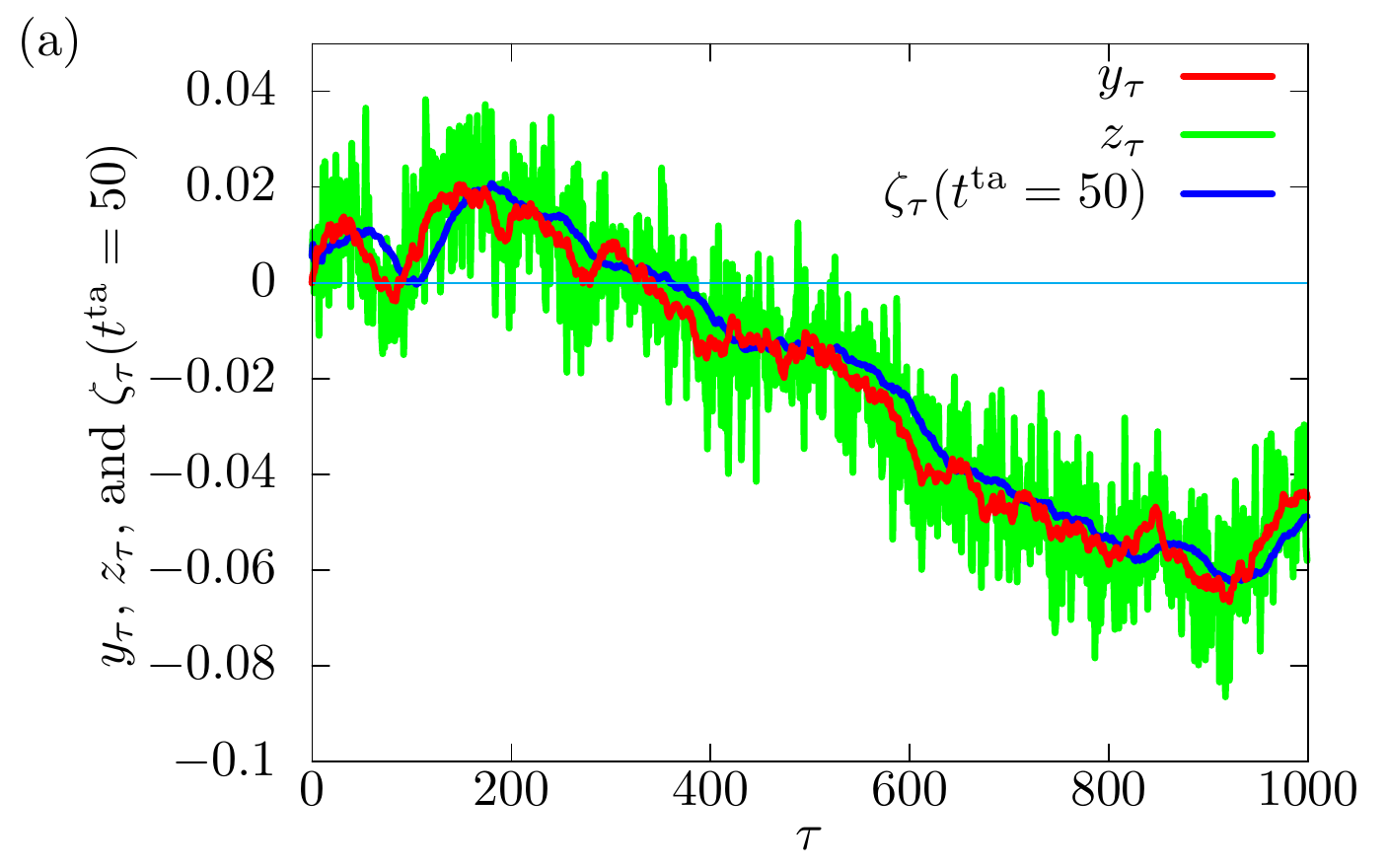}
\includegraphics[scale=0.50]{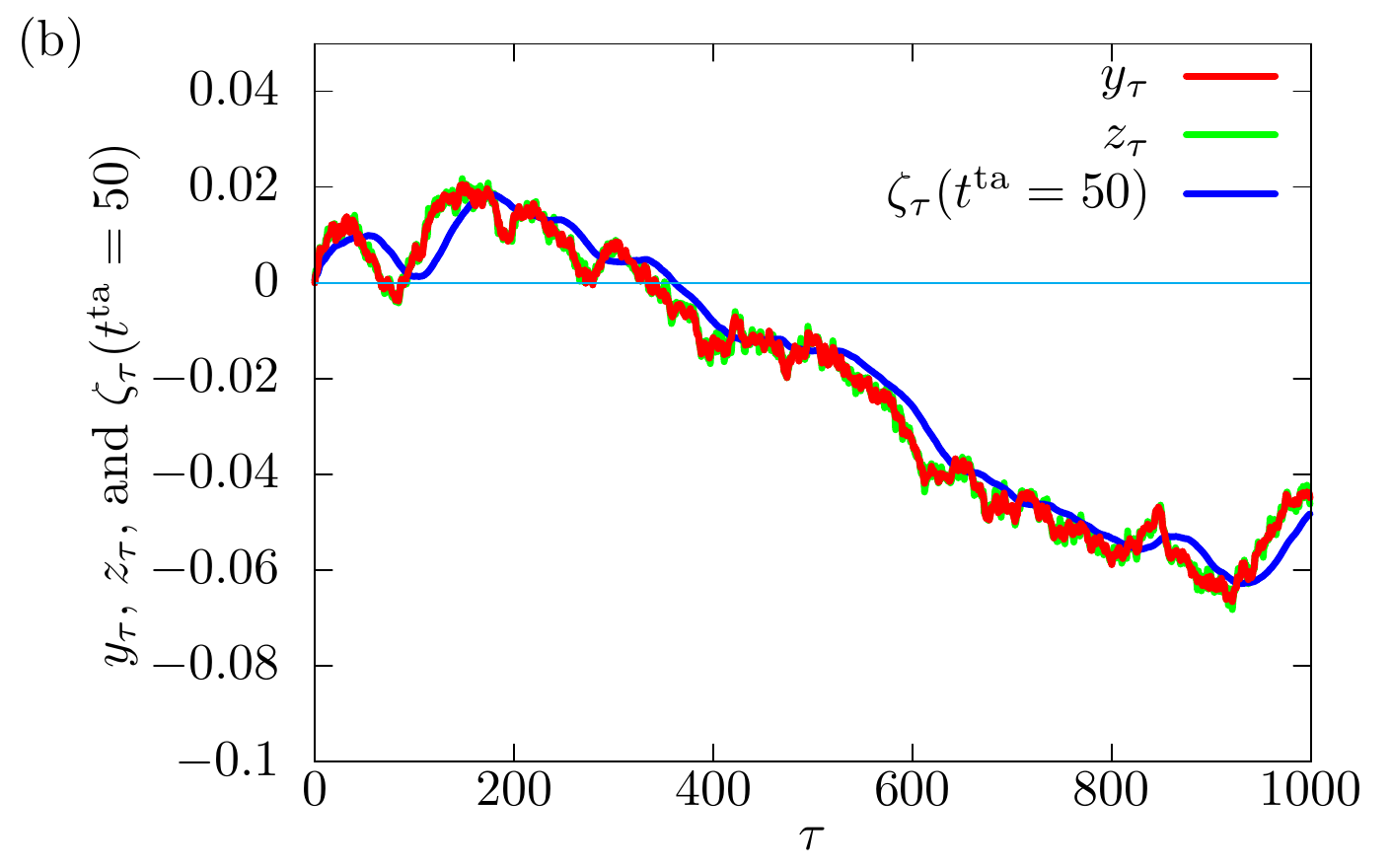}
\caption{Dynamics of $y_\tau$ (red line), $z_\tau$ (green line), and $\zeta_\tau (t^\mathrm{ta} = 50)$ (blue line) in the cases of (a) $\sigma_z = 0.010$ and (b) $\sigma_z = 0.0010$.}
\label{eq_numerical_sim_001_001}
\end{figure}

\begin{figure}[tb]
\centering
\includegraphics[scale=0.50]{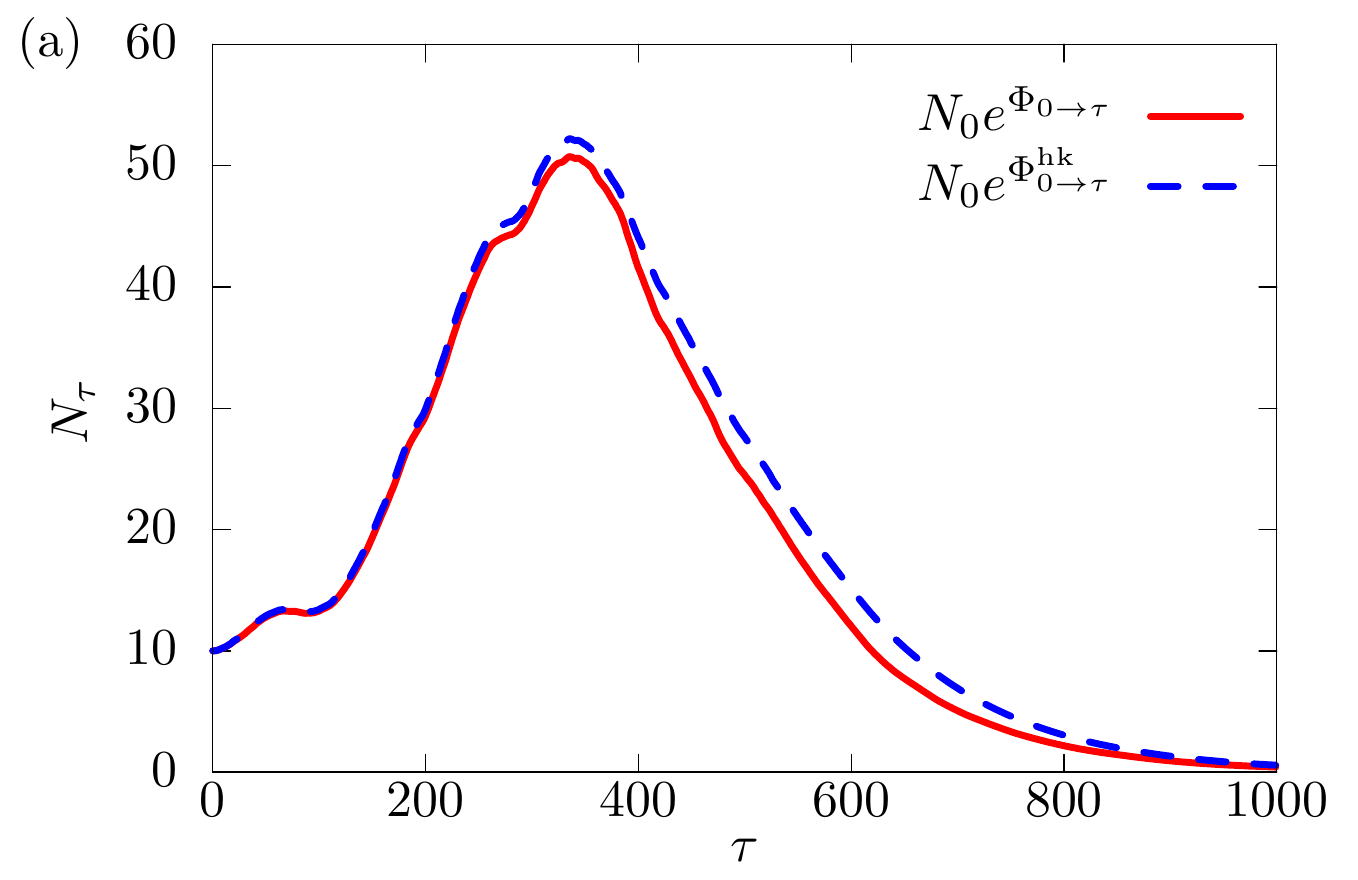}
\includegraphics[scale=0.50]{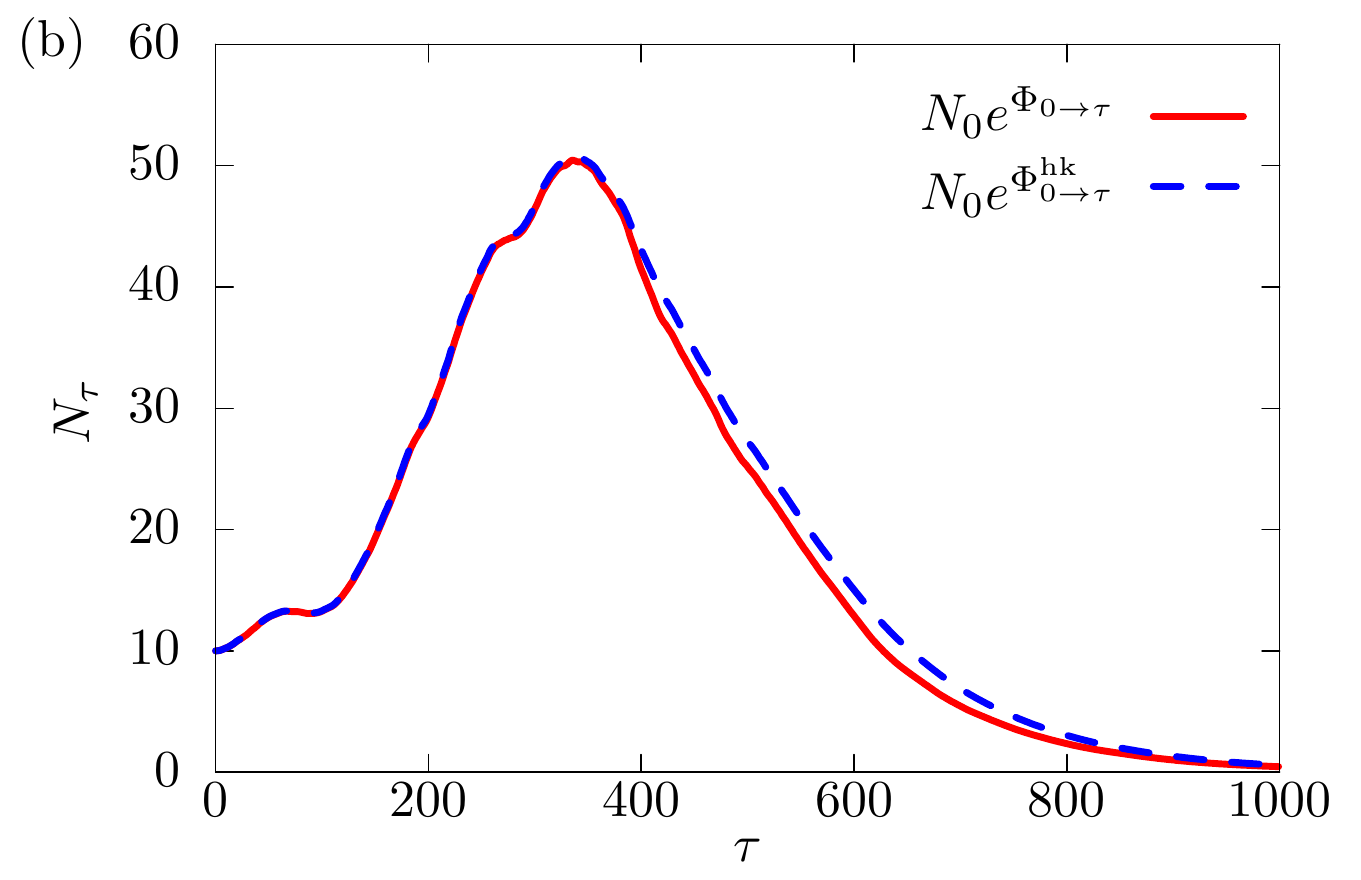}
\caption{Total number of organisms $N_0 e^{\Phi_{0 \to \tau}}$ (red solid line) and housekeeping number of organisms $N_0 e^{\Phi_{0 \to \tau}^\mathrm{hk}}$ (blue dashed line) for (a) $\sigma_z = 0.010$ and (b) $\sigma_z = 0.0010$ in the case of the naive strategy.}
\label{eq_numerical_sim_002_001}
\end{figure}

\begin{figure}[tb]
\centering
\includegraphics[scale=0.50]{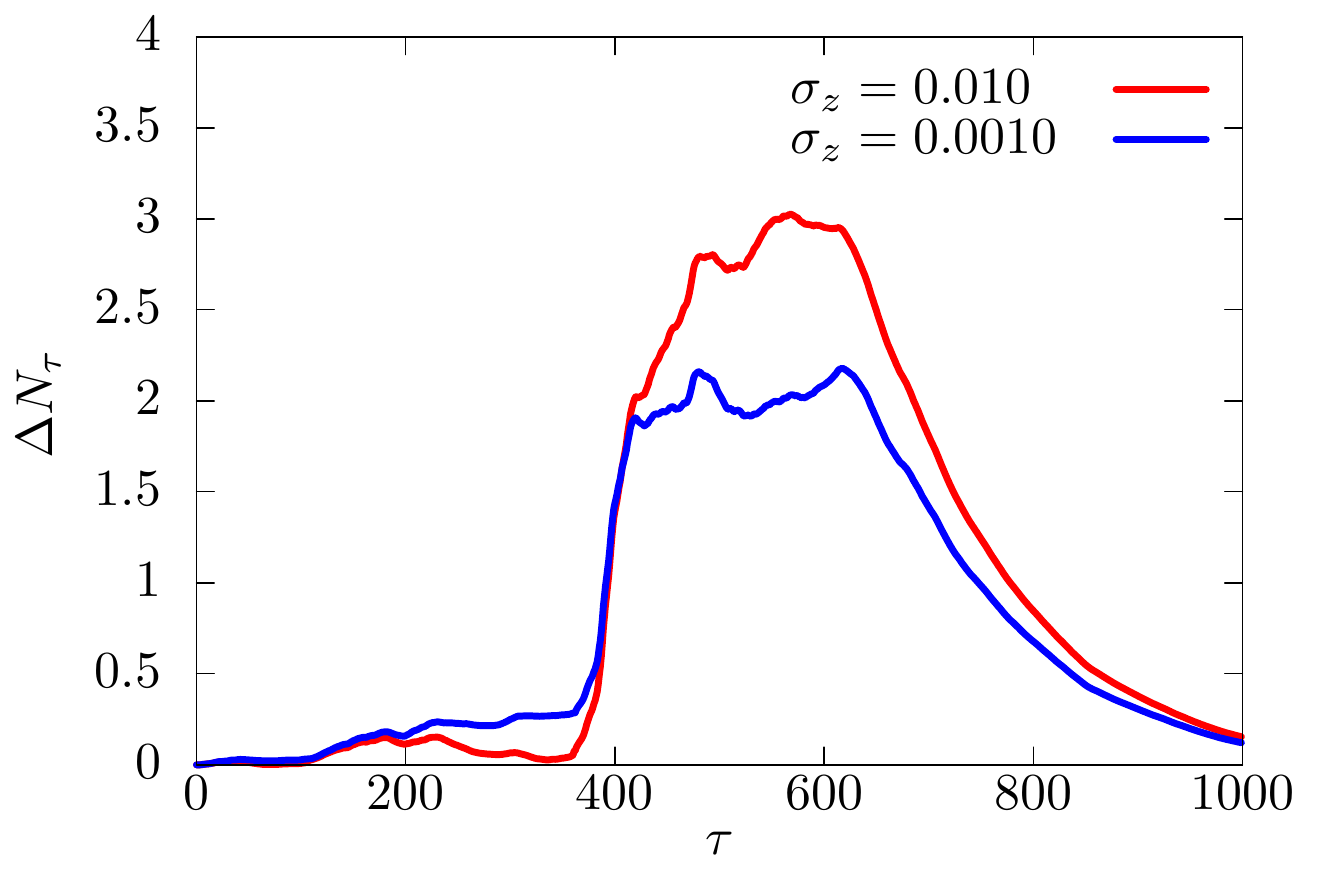}
\caption{{Gap between the number of organisms and its bound in the case of the naive strategy, $\Delta N_\tau$,} for $\sigma_z = 0.010$ and $\sigma_z = 0.0010$.}
\label{eq_numerical_sim_002_011}
\end{figure}

In Fig.~\ref{eq_numerical_sim_001_001}, we show the dynamics of $y_\tau$ and $z_\tau$ in red and green lines, respectively.
Depending on $\sigma_z$, green lines in Figs.~\ref{eq_numerical_sim_001_001}(a) and (b) are quite different.
In Figs.~\ref{eq_numerical_sim_002_001}(a) and (b), we show $N_0 e^{\Phi_{0 \to t}}$ and $N_0 e^{\Phi_{0 \to t}^\mathrm{hk}}$.
We set $\sigma_z = 0.010$ and $\sigma_z = 0.0010$ in Figs.~\ref{eq_numerical_sim_002_001}(a) and (b), respectively.
In both cases, $N_0 e^{\Phi_{0 \to t}}$ is bounded by $N_0 e^{\Phi_{0 \to t}^\mathrm{hk}}$, and thus this numerical result supports Eq.~\eqref{eq_inequality_checked_001_002}.
Note that this process is sufficiently long; thus the pseudo-entropy difference is negligible.
In Fig.~\ref{eq_numerical_sim_002_011}, we plot {the gap between the number of organisms and its bound, $\Delta N_\tau \coloneqq N_0 e^{\Phi_{0 \to \tau}^\mathrm{hk}} - N_0 e^{\Phi_{0 \to \tau}}$, for both cases.}
This figure implies that in the case of small noise on information, {$\Delta N_\tau$ gets small}.
Thus, it also tells us that, to attain the bound, accurate information is critically important.

\subsection{Numerical simulation II} \label{sec_numerical_simulation_001_002}

We focus on the time-averaging strategy.
We again use the numerical setup shown above.
For the time-averaging strategy, we use the moving average represented by Eq.~\eqref{eq_moving_average_001_001} and put $t^\mathrm{ta} = 50$.
Blue lines in Fig.~\ref{eq_numerical_sim_001_001} denote $\zeta_\tau (t^\mathrm{ta} = 50)$.
We denote, by $\Phi_{0 \to t}^\mathrm{ta, hk}$, {housekeeping growth, Eq.~\eqref{eq_def_housekeeping_growth_001_001},} in the case of the time-averaging strategy.

\begin{figure}[tb]
\centering
\includegraphics[scale=0.50]{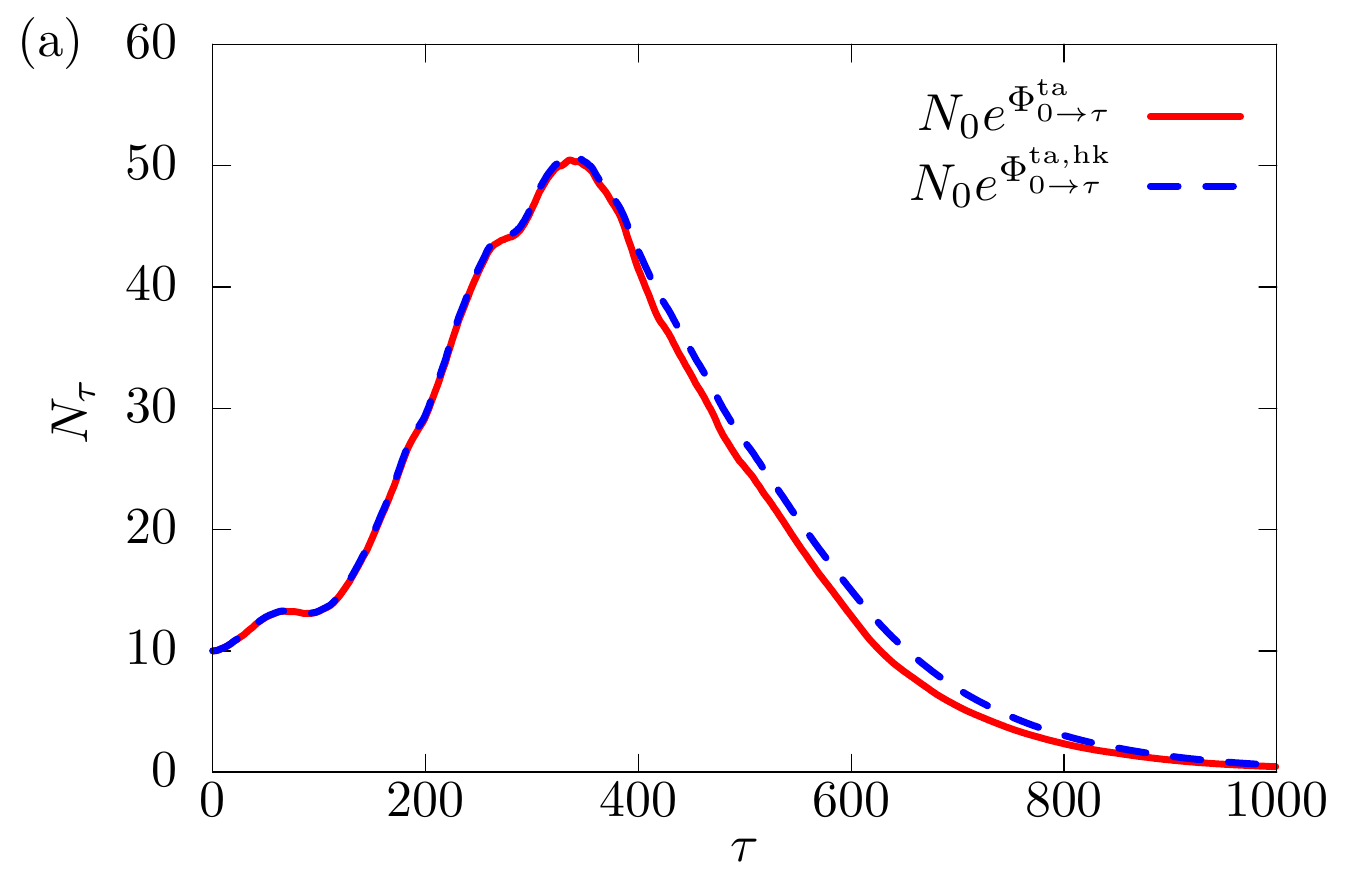}
\includegraphics[scale=0.50]{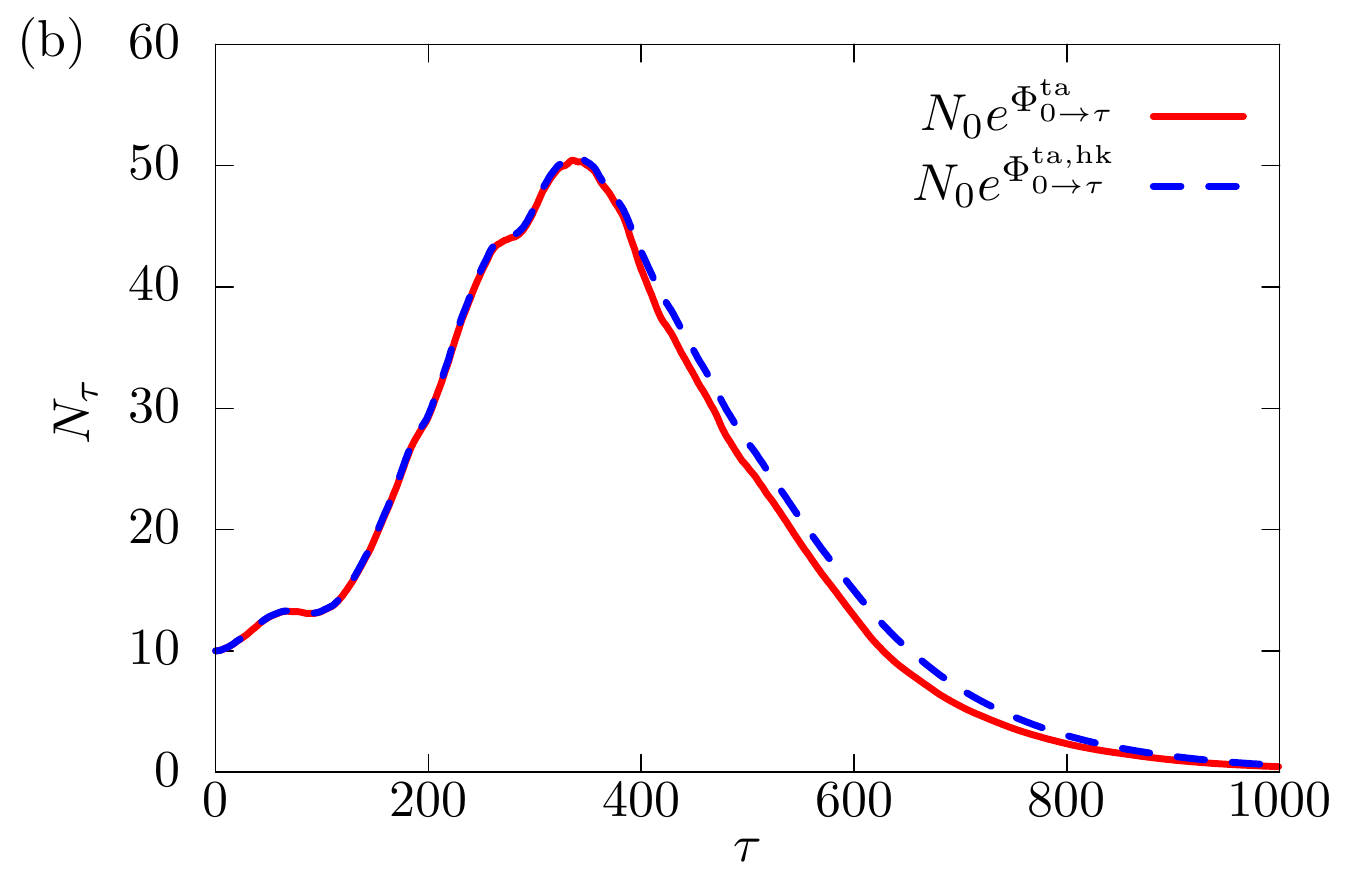}
\caption{Total number of organisms $N_0 e^{\Phi_{0 \to \tau}^\mathrm{ta}}$ (red solid line) and housekeeping number of organisms $N_0 e^{\Phi_{0 \to \tau}^\mathrm{ta, hk}}$ (blue dashed line) for (a) $\sigma_z = 0.010$ and (b) $\sigma_z = 0.0010$ in the case of the time-averaging strategy.}
\label{numerical-sim-02-02}
\end{figure}

\begin{figure}[tb]
\centering
\includegraphics[scale=0.50]{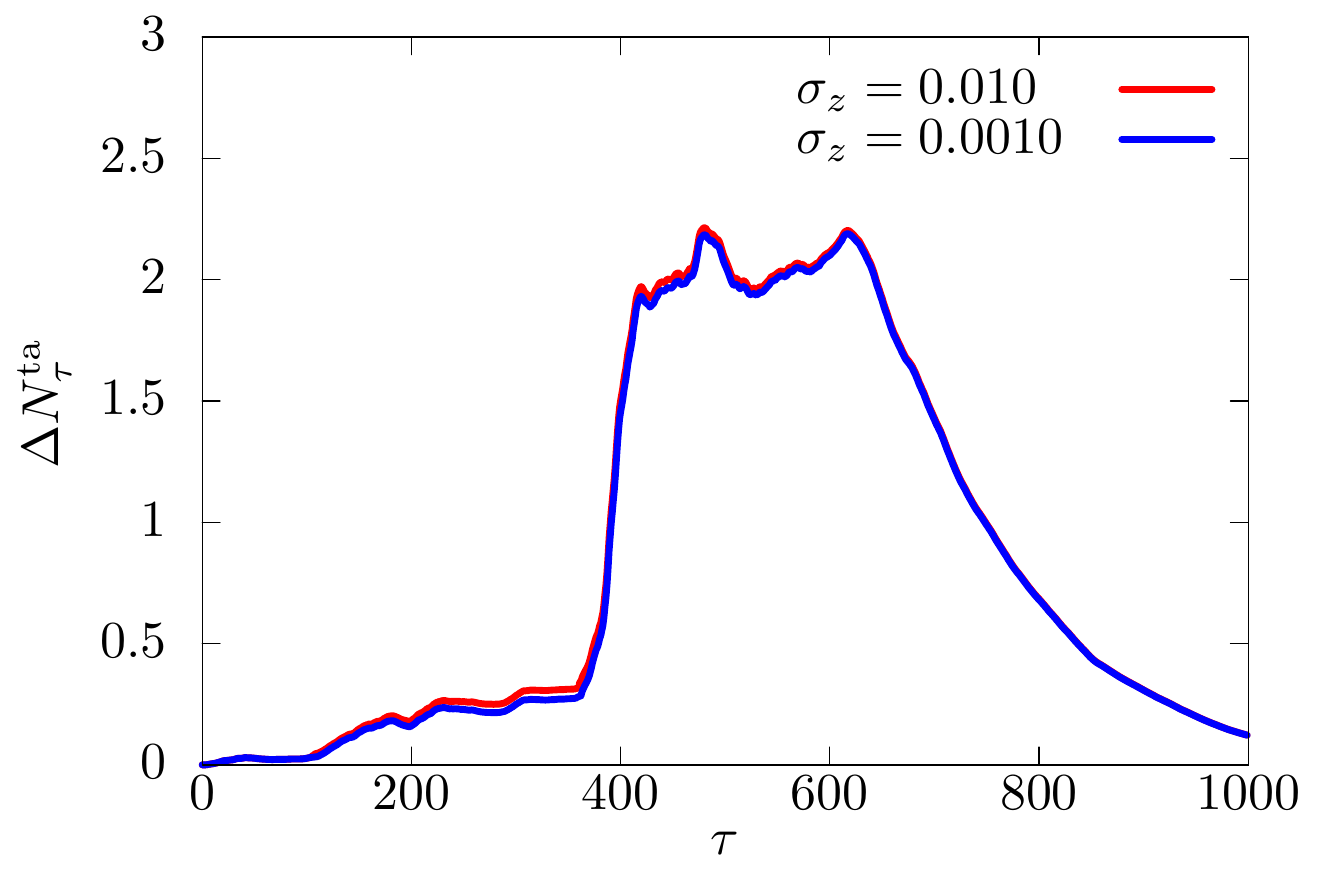}
\caption{{Gap between the number of organisms and its bound in the case of the time-averaging strategy, $\Delta N_\tau^\mathrm{ta}$,} for $\sigma_z = 0.010$ and $\sigma_z = 0.0010$.}
\label{numerical-sim-02-12}
\end{figure}

Similarly to the above numerical simulation, we added larger Gaussian noise with $\sigma_z = 0.010$ on $z_\tau$ in Fig.~\ref{numerical-sim-02-02}(a) and smaller Gaussin noise with $\sigma_z = 0.0010$ on $z_\tau$ in Fig.~\ref{numerical-sim-02-02}(b).
Figures~\ref{numerical-sim-02-02}(a) and (b) show that Eq.~\eqref{eq_inequality_checked_001_002} holds for the time-averaging strategy.
In Fig.~\ref{numerical-sim-02-12}, we plot {the gap between the number of organisms and its bound, $\Delta N_\tau^\mathrm{ta} \coloneqq N_0 e^{\Phi_{0 \to \tau}^\mathrm{ta, hk}} - N_0 e^{\Phi_{0 \to \tau}^\mathrm{ta}}$,} for both cases.
This figure shows that $\Delta N_\tau^\mathrm{ta}$ is almost the same {for both cases} and insensitive to the magnitude of noise.
This implies that the time-averaging strategy is robust and useful under noisy situations.
Note that, to implement the time-averaging strategy, organisms must have memories.

% !TEX root = ./paper_population_dynamics_000_001_main.tex

\section{Discussions} \label{sec_discussions_001}

{This section discusses the findings of this paper.}
{We first explain the novelty of the Clausius inequality, Eq.~\eqref{eq_Clausius_inequality_001_001}, by comparing it with a trivial bound.
Then we explain the gap between them by using perturbation theory on eigenvalues.}
{To clarify the novelty of our findings, we also mention the relationship between the Clausius inequality, Eq.~\eqref{eq_Clausius_inequality_001_001}, and the Clausius inequality found in Ref.~\cite{Sughiyama_002}.
Finally, we explain that a similar relationship also holds for the expected values.}

\subsection{Novelty of the Clausius inequality}

{In short, the novelty of the Clausius inequality, Eq.~\eqref{eq_Clausius_inequality_001_001}, is that excess growth, Eq.~\eqref{eq_def_excess_growth_001_001}, is bounded by the difference between the path-independent quantities, Eq.~\eqref{eq_def_pseudo-entropy_001_001}, which we call the pseudo-entropy, at the beginning and end of a given process.
We elaborate on the novelty of the Clausius inequality, Eq.~\eqref{eq_Clausius_inequality_001_001}, by using Eq.\eqref{eq_inequality_checked_001_001} because it provides us with a slightly different perspective.
Let us introduce $\mu_y^\mathrm{max} \coloneqq \max_x \mu_y (x)$, which is the maximum duplication rate when $y$ is given.
Since $\phi_{y, z}^\mathrm{st} \le \mu_y^\mathrm{max}$,
we have a following trivial bound:
\begin{align}
  \Phi_{0 \to t} &\le \Phi_{0 \to t}^\mathrm{max} + S_{y_t, z_t} - S_{y_0, z_0}. \label{eq_Clausius_inequality_003_001}
\end{align}
where
\begin{align}
  \Phi_{0 \to t}^\mathrm{max} &\coloneqq \sum_{\tau = 1}^t \mu_{y_\tau}^\mathrm{max}. \label{eq_Phi_max_001_001}
\end{align}
If $\{ y_\tau \}_{\tau = 0, 1, \dots, t}$ are given, then Eq.~\eqref{eq_Clausius_inequality_003_001} becomes a good bound because organisms can always select the phenotype that has the largest duplication rate for given $y_\tau$.
However, in a practical setup, organisms cannot get $y_\tau$ directly.
Thus, the Clausius inequality, Eq.~\eqref{eq_Clausius_inequality_001_001}, provides us with a tighter bound than Eq.~\eqref{eq_Clausius_inequality_003_001} for a practical setup.}

\subsection{Perturbation analysis of the Clausius equality and inequality}

By using the perturbation theory on eigenvalues, we have
\begin{align}
\phi_{y, z}^\mathrm{st} &\approx \mu_y^\mathrm{max} + [\vec{v}_{y, z}^\mathrm{max}]^\intercal \hat{M}_y (\hat{T}_{z} - \hat{1}_n) \vec{v}_{y, z}^\mathrm{max} \\
&= \mu_y^\mathrm{max} + \mu_y^\mathrm{max} [\vec{v}_{y, z}^\mathrm{max}]^\intercal (\hat{T}_{z} - \hat{1}_n) \vec{v}_{y, z}^\mathrm{max}, \label{eq_perturbation_eigenvalues_001_001}
\end{align}
where $\hat{1}_n$ is the $n \times n$ identity matrix and $\vec{v}_{y, z}^\mathrm{max}$ satisfies $\hat{M}_y \vec{v}_{y, z}^\mathrm{max} = \mu_y^\mathrm{max} \vec{v}_{y, z}^\mathrm{max}$ and $\| \vec{v}_{y, z}^\mathrm{max} \|_1 = \| \vec{v}_{y, z}^\mathrm{max} \|_2 = 1$.
Note that the above approximation is valid when $T_{z} (x| x') \approx \delta_{x, x'}$.
Here, $\delta_{x, x'}$ is the Kronecker delta function.

As we explained, {the Clausius inequality, Eq.~\eqref{eq_Clausius_inequality_001_001},} is a bound when we adopt a $z$-dependent transition matrix $T_{z} (x| x')$, and Eq.~\eqref{eq_Clausius_inequality_003_001} is a looser bound defined by the maximum {duplication} growth rate.
The gap between Eqs.\eqref{eq_inequality_checked_001_001} and \eqref{eq_Clausius_inequality_003_001} to the first-order perturbation is the summation of the second term of Eq.~\eqref{eq_perturbation_eigenvalues_001_001} for $\tau = 1, 2, \dots, t$, and it clearly shows that, by considering {the phenotype-switching rate} described by $T_{z} (x| x')$, we can obtain a tighter bound for the growth rate.

\subsection{Relation with the case without information}

Until now, we have considered the case in which we can utilize information on environments.
Let us consider organisms that cannot sense their environments, that is, the case of a fixed transition matrix.
Then, {the Clausius inequality, Eq.~\eqref{eq_Clausius_inequality_001_001}} becomes
\begin{align}
  \Phi_{0 \to t}^{\mathrm{ex}, \text{(w/o info)}} &\le S_{y_t, c} - S_{y_0, c},
\end{align}
where $\Phi_{0 \to t}^{\mathrm{ex}, \text{(w/o info)}}$ is {excess growth, Eq.~\eqref{eq_def_excess_growth_001_001}, in the case of} $C_{0 \to t} \coloneqq \{ c \}_{\tau = 1}^t$ and $c$ is a constant that does not depend on $\tau$.
This result is equivalent with Ref.~\cite{Sughiyama_002}.

\subsection{Expectation of the Clausius inequality}

{The Clausius inequality, Eq.~\eqref{eq_Clausius_inequality_001_001}} is the inequality for each realization of ${Y_{0 \to t}}$ and ${Z_{0 \to t}}$; so its expected value is easily computed as
\begin{align}
\langle \Phi_{0 \to t} \rangle_{p ({Y_{0 \to t}}, {Z_{0 \to t}})} &\le \langle \Phi_{0 \to t}^\mathrm{hk} \rangle_{p ({Y_{0 \to t}}, {Z_{0 \to t}})} + \langle S_{y_t, z_t} \rangle_{p (y_t, z_t)} - \langle S_{y_0, z_0} \rangle_{p (y_0, z_0)}. \label{eq_Clausius_inequality_002_004}
\end{align}
Here $p (y_\tau, z_\tau)$ is the distribution of $y_\tau$ and $z_\tau$.
{From the viewpoint of experimental verifications, Eq.~\eqref{eq_Clausius_inequality_002_004} may be more convenient than the Clausius inequality, Eq.~\eqref{eq_Clausius_inequality_001_001}.}

% !TEX root = ./paper_population_dynamics_000_001_main.tex

\section{Conclusions} \label{sec_conclusions_001}

{In this paper, we have constructed an SST structure with side information in population dynamics and derived novel relations, the Clausius equality and inequality.
First, we defined population growth of organisms and divided it into two parts: housekeeping growth and excess growth.
Then we showed that excess growth is bounded by the difference of path-independent quantities, which we call the pseudo-entropy, at the beginning and end.
We also performed numerical simulations to confirm our findings with several different setups.
Finally, we explained the relationship between our findings and previous ones and discussed their novelties.
To the best of our knowledge, this paper has found an SST structure with side information in interdisciplinary sciences for the first time.}

% !TEX root = ./paper_population_dynamics_000_001_main.tex

\begin{acknowledgments}
  H.M. thanks Kiyoshi Kanazawa, Yuki Sughiyama, and Tetsuya J. Kobayashi for fruitful discussions.
\end{acknowledgments}

\appendix

% \input{paper_population_dynamics_071_001}
% !TEX root = ./paper_population_dynamics_000_001_main.tex

\section{Non-negativity of the symmetrized divergence} \label{app_sec_nonnegativity_divergence_001_001}

We show that $D_{0 \to t}^\mathrm{sym}$ in Eq.~\eqref{eq_def_divergence_sym_001_001} is non-negative.
The proof is as follows:
\begin{align}
D_{0 \to t}^\mathrm{sym} &= \int \mathcal{D} X_{0 \to t} \, p_\mathrm{b} (X_{0 \to t}| Y_{0 \to t}, Z_{0 \to t}) \ln \frac{p_\mathrm{b} (X_{0 \to t}| Y_{0 \to t}, Z_{0 \to t})}{\big[ p_\mathrm{r} (X_{0 \to t}| Y_{0 \to t}, Z_{0 \to t}) p_\mathrm{dr} (X_{0 \to t}^\mathrm{r}| Y_{0 \to t}^\mathrm{r}, Z_{0 \to t}^\mathrm{r}) \big]^{\frac{1}{2}}} \\
&= \int \mathcal{D} X_{0 \to t} \, p_\mathrm{b} (X_{0 \to t}| Y_{0 \to t}, Z_{0 \to t}) \ \nonumber \\
& \quad \times \bigg( \ln p_\mathrm{b} (X_{0 \to t}| Y_{0 \to t}, Z_{0 \to t}) - \frac{1}{2} \ln p_\mathrm{r} (X_{0 \to t}| Y_{0 \to t}, Z_{0 \to t}) - \frac{1}{2} \ln p_\mathrm{dr} (X_{0 \to t}^\mathrm{r}| Y_{0 \to t}^\mathrm{r}, Z_{0 \to t}^\mathrm{r}) \bigg) \\
&= - \frac{1}{2} \int \mathcal{D} X_{0 \to t} \, p_\mathrm{b} (X_{0 \to t}| Y_{0 \to t}, Z_{0 \to t}) \ln \frac{p_\mathrm{r} (X_{0 \to t}| Y_{0 \to t}, Z_{0 \to t})}{p_\mathrm{b} (X_{0 \to t}| Y_{0 \to t}, Z_{0 \to t})} \nonumber \\
& \quad - \int \frac{1}{2} \mathcal{D} X_{0 \to t} \, p_\mathrm{b} (X_{0 \to t}| Y_{0 \to t}, Z_{0 \to t}) \ln \frac{p_\mathrm{dr} (X_{0 \to t}^\mathrm{r}| Y_{0 \to t}^\mathrm{r}, Z_{0 \to t}^\mathrm{r})}{p_\mathrm{b} (X_{0 \to t}| Y_{0 \to t}, Z_{0 \to t})} \\
&\ge - \frac{1}{2} \int \mathcal{D} X_{0 \to t} \, p_\mathrm{r} (X_{0 \to t}| Y_{0 \to t}, Z_{0 \to t}) - \frac{1}{2} \int \mathcal{D} X_{0 \to t} \, p_\mathrm{dr} (X_{0 \to t}| Y_{0 \to t}, Z_{0 \to t}) \\
& = 0.
\end{align}
Note that we have used Jensen's inequality.

%

% !TEX root = ./paper_population_dynamics_000_001_main.tex

% \bibliographystyle{apsrev4-1}
% \bibliographystyle{unsrt}
\bibliography{paper_population_dynamics_999_001}

\end{document}